\documentclass[10pt]{article}

\usepackage{amsmath,amssymb} 
\usepackage{mathrsfs}
\usepackage{graphicx} 
\usepackage{color}
\usepackage{mathtools}
\usepackage{float}
\usepackage{bbm}
\usepackage{appendix}
\usepackage[table,xcdraw]{xcolor}

\usepackage[caption=false]{subfig}
\def \R{{\mathbb{R}}}

\newtheorem{remark}{Remark}

\title{The dissipative Generalized Hydrodynamic equations and their numerical solution} 

\author{F. M{\o}ller$^1$, N. Besse$^2$, I.E. Mazets$^{1,3}$, H.P. Stimming$^3$, N.J. Mauser$^3$} 

\date{\small
  \begin{flushleft}
    $^1$ Vienna Center for Quantum Science and Technology (VCQ), Atominstitut, TU Wien, 1020 Vienna, Austria\\ 
    $^2$ Observatoire de la C\^ote d'Azur, Univ. C\^ote d'Azur, 
    06300 Nice, Cedex 4, France\\
    $^3$  Research platform MMM "Mathematics-Magnetism-Materials" c/o Fak. Mathematik, Univ. Wien, 1090 Vienna, Austria
  \end{flushleft}
}

\begin{document}   

\maketitle

\begin{abstract} 
  ``Generalized Hydrodynamics'' (GHD) stands for a model that describes one-dimensional
  \textit{integrable} systems in quantum physics, such as ultra-cold atoms or spin chains.
  Mathematically, GHD corresponds to nonlinear equations of kinetic type,
  where the main unknown, a statistical distribution function
  $f(t,z,\theta)$, lives in a phase space which is constituted by a one-dimensional position variable $z$,
  and a one-dimensional "kinetic" variable $\theta$, actually a wave-vector, called ``rapidity''.
  Two key features of GHD equations are first a non-local and nonlinear coupling in the advection term,
  and second an infinite set of conserved quantities, which prevent the system from thermalizing. 
  To go beyond this, we consider the dissipative GHD equations, which are obtained by supplementing
  the right-hand side of the GHD equations with a non-local and nonlinear diffusion operator or
  a Boltzmann-type collision integral.
  In this paper, we deal with new high-order numerical methods to efficiently solve these kinetic equations.
  In particular, we devise novel backward semi-Lagrangian methods for solving the advective part
  (the so-called Vlasov equation) by using a high-order time-Taylor series expansion for the advection fields,
  whose successive time derivatives are obtained by a recursive procedure. This high-order temporal approximation
  of the advection fields are used
  to design new implicit/explicit Runge--Kutta semi-Lagrangian methods, which are compared to Adams--Moulton
  semi-Lagrangian schemes. For solving the source terms, constituted by the diffusion and collision operators,
  we use and compare different numerical methods of the literature.
\end{abstract} 


\medskip

Generalized HydroDynamics,
Backward Semi Lagrangian methods,
High order Runge Kutta methods.

\section{Introduction}

Over the last decades, significant advances in experimentally realizing and manipulating quantum many-body systems have been made. 
In particular, the platform of ultracold atoms has demonstrated great success, owing to its versatility; by controlling the trapping geometry or interactions of the atoms, the effective Hamiltonian of the system can be tuned to follow various condensed matter or field theory models~\cite{RevModPhys.80.885, RevModPhys.83.1405, RevModPhys.80.1215}.
Hence, ultracold gases are sometimes referred to as 'analog quantum simulators'~\cite{Bloch2012, doi:10.1126/science.aal3837}.

However, the behavior of such interacting quantum many-body systems is notoriously complex, thus leading to a high demand for theoretical methods capable of simulating their dynamics.
Particularly successful theoretical developments include, among others, the Density Matrix Renormalization Group~\cite{Daley_2004, RevModPhys.77.259, SCHOLLWOCK201196} and quantum Monte Carlo methods~\cite{ACIOLI199775, RevModPhys.73.33, Pollet_2012}.
Alternative to treating microscopic interactions, emergent properties of interacting quantum many-body system can often be described in terms of quasi-particles representing collective degrees of freedom.
The concept of quasi-particles was originally developed to study the transport of electrons in a solid~\cite{Landau1933}, but has since been employed to describe a number of phenomena, such as colossal magnetoresistance~\cite{Mannella2005} and superconductivity~\cite{RevModPhys.78.17}.

In \textit{integrable} systems, quasi-particles descriptions are especially powerful, as they provide an exact solution following the Bethe Ansatz~\cite{Bethe1931, lieb2013mathematical, gaudin2014bethe, korepin_bogoliubov_izergin_1993}.
Such systems include, but are not limited to, repulsively interacting Bose~\cite{LL1, LL2} and Fermi~\cite{PhysRevLett.19.1312, GAUDIN196755} gases and the anisotropic XXZ Heisenberg spin chain~\cite{PhysRev.112.309}.
Integrable systems share a number of key properties; they are one-dimensional, all associated multi-particle scattering matrices factorize into two-particle matrices, and they obey an infinite number of conservation laws.
Further, by virtue of their restricted geometry, the systems are intrinsically strongly correlated and exhibit phenomena not found in higher-dimensional system, such as the absence of thermalization~\cite{PhysRevLett.103.100403, PhysRevLett.110.257203, doi:10.1126/science.1257026}.
The resulting dynamics and relaxation properties are thus rather particular (see the special issue~\cite{Calabrese_2016}), as is perhaps best demonstrated in the seminal quantum Newton's cradle experiment~\cite{kinoshita2006quantum}.

Recently, the development of the theory of "Generalized Hydrodynamics" (GHD)~\cite{castro2016emergent, bertini2016transport} has drastically simplified the study of quantum many-body dynamics in integrable system by formulating the conservation laws as a kinetic equation for the quasi-particles.
The theory derives it name from the generalized Gibbs ensemble~\cite{PhysRevLett.98.050405, Rigol2008}; these statistical ensembles describe the non-thermal, maximum-entropy steady states of integrable models.
Following its inception, a large number of extensions have been made to GHD (see the special issue~\cite{Bastianello_2022}), treating the spreading of entanglement and correlations~\cite{Alba_2021}, transport properties~\cite{DeNardis_2022, Bulchandani_2021} and perturbations to integrability~\cite{PhysRevLett.126.090602, Bastianello_2021}.
The validity of GHD and some of its extensions have been demonstrated following observations from several experiments with (quasi) one-dimensional Bose gases~\cite{schemmer2019generalized, malvania2020generalized, PhysRevLett.126.090602, PhysRevX.12.041032} (see also the review~\cite{Bouchoule_2022}).


However, despite its significance and impact ~\cite{10.21468/SciPostPhys.8.3.041, 10.21468/SciPostPhysCore.3.2.016, PhysRevLett.120.045301, https://doi.org/10.48550/arxiv.2208.06614, PhysRevLett.119.195301, PhysRevLett.124.140603, https://doi.org/10.48550/arxiv.2205.15871}, no systematic approach to develop and compare numerical methods has been done in relation to GHD, with most studies employing only low-order schemes~\cite{PhysRevLett.119.220604, PhysRevLett.123.130602, PhysRevB.97.045407, PhysRevB.103.L060302, PhysRevLett.125.240604}.
For advancing the use of GHD, especially for numerical modeling of experiments, it is very useful to further develop and test numerical techniques for solving the associated kinetic equations.
In this work, we develop a set of novel backward semi-Lagrangian (BSL) schemes of high-order (implicit/explicit Runge--Kutta semi-Lagrangian schemes), which give very accurate results proved by comparison with other high-order semi-Lagrangian schemes (Adams--Moulton). 
Our results represent the highest numerical precision solution of the GHD equation so far.
Further, we present several versions of BSL hybrid schemes for solving the GHD equation in the presence of diffusion (higher order corrections to the typical Euler scale formulation of GHD) or non-integrable perturbations (accounted for by a Boltzmann-type collision integral).
Of the latter we here treat the effect of highly energetic collisions in a Bose gas, which can lead to 3D excitations of the atoms in the trapping potential used to confine the gas.
However, the methods presented can be generalized to a number of other relevant perturbations~\cite{Bastianello_2021, PhysRevB.101.180302, PhysRevLett.127.130601, https://doi.org/10.48550/arxiv.2205.06492} in a straightforward manner.\\

This paper is organized as follows: 
In Section~\ref{s:GHD}, we introduce the theory of Generalized Hydrodynamics (GHD) and present the key equations, namely the advective and dissipative GHD equations.
Of the latter, we focus on two particular dissipative source terms: the diffusion operator and the Boltzmann-type collision integral.
Next, in Section~\ref{s:ATDAF}, we derive exact time-derivatives of the GHD advection fields, while in Section~\ref{s:dressing} we discuss how to solve the dressing equation ubiquitous in GHD.
In Section~\ref{s:BSL}, we recall the backward semi-Lagrangian method, followed, in Section~\ref{s:schemes}, by the presentation of various high-order time discretization schemes.
In particular, we develop novel implicit/explicit Runge--Kutta methods.
In Section~\ref{s:benchmarks}, we conduct a number of numerical benchmarks, testing the accuracy of BSL schemes for solving the advective GHD equation and BSL-hybrid schemes for solving the dissipative GHD equation.
Finally, in Section~\ref{s:conclusion}, we present some concluding remarks.

\section{The GHD equations} \label{s:GHD}
Here, we present the GHD equations in two steps. In Section~\ref{ss:GHDAE}, we first
describe the GHD advection equation, which can be seen as a collisionless kinetic equation of
Vlasov type with non-local and nonlinear advection fields. Then, in Section~\ref{ss:ECD},
we present two source terms for the right-hand side of the GHD equation, resulting in a advective-dissipative equation.
The first source term is a diffusion operator (see Section~\ref{sss:Dop}), while the second one is
a Boltzmann-type collision operator (see Section~\ref{sss:Colop}). Both operators are
are non-local and nonlinear, but their dissipation mechanisms and mathematical structures are very different.

\subsection{The GHD advection equation}
\label{ss:GHDAE}

By virtue of the infinite number of conservation laws found in integrable systems, their emergent hydrodynamics consist, in principle, of an infinite set of continuity equations.
Generalized Hydrodynamics employs the so-called "thermodynamic Bethe Ansatz"~\cite{takahashi_1999} to formulate all the conservation laws as a single kinetic equation.
Following the thermodynamic Bethe Ansatz, the local equilibrium state of an integrable systems is fully characterized by a density of \textit{quasi-particles} $\rho_\mathrm{p}$.
The quasi-particles represent collective excitations of the system and are uniquely labelled by their wavevector $\theta$, the rapidity, which in the thermodynamic limit becomes a continuous parameter. Note that with $\theta$ having dimension 1/lenght, $p = \hbar \theta$ is a momentum.
Assuming small variations between neighbouring mesoscopic regions of the system, a single position-dependent quasi-particle density describes the macrostate of the entire system.
Large scale dynamics can then be viewed as the propagation of quasi-particles between different regions of the system.
The mean propagation velocity of a given quasi-particle, dubbed the \textit{effective velocity}, is highly dependent on interactions with other quasi-particles at the same point in space.
At the lowest order (Euler scale), where all currents depend only on the local densities of conserved quantities, the propagation of the quasi-particles is purely ballistic.
Expressions for the currents were first presented in \cite{castro2016emergent, bertini2016transport} and later more rigorously proven in \cite{Cubero_2021, Borsi_2021, El_2021, Buca_2021}.
Including first-order positional derivatives of the densities in the current functions results in diffusive corrections to the quasi-particle propagation~\cite{PhysRevLett.121.160603, NBD19}.

The most convenient formulation of the GHD equation is not in terms of the quasi-particle density $\rho_{\mathrm{p}}$ but rather the filling function $f$.
The filling function is non-negative and can be seen as a statistical distribution function of particle occupation rate in the phase space, encoding the fraction of possible momentum states occupied by the particles.
In fact, the functions $\rho_{\mathrm{p}}$ and $f$ can be considered as two different sets of state coordinates, which each can be used to fully characterize the local equilibrium state~\cite{takahashi_1999}.
They are related by the fundamental identity \eqref{eqn:frhoprhos} specified below. 
Furthermore, each rapidity component of the filling function can be identified as a fluid mode~\cite{castro2016emergent, 10.21468/SciPostPhysLectNotes.18}.
Following GHD, the infinite set of advection equations for the fluid modes can be written as a single, two-dimensional advection equation for the filling function $f=f(t,z,\theta)$, namely
\begin{equation}
  \partial_{t} f+v^{\mathrm{eff}} \: \partial_{z} f+a^{\mathrm{eff}} \: \partial_{\theta} f =0,
    \label{eq:GHD_equation}
\end{equation}
where $t\in \R_+$,  $z\in \R$,  and $\theta \in \R$, represent, respectively, the time, position and momentum (rapidity) variables \cite{castro2016emergent, bertini2016transport}.
The advection equation \eqref{eq:GHD_equation}, which is the main equation of the GHD equations, must be supplemented with an initial condition, $f(0,z,\theta)=f_0(z,\theta)$.
The effective velocity $v^{\mathrm{eff}}=v^{\mathrm{eff}}(t,z,\theta)$ is a non-local and nonlinear expression in terms of $f$.
Similarly, the effective acceleration $a^{\mathrm{eff}}=a^{\mathrm{eff}}(t,z,\theta)$, accounting for inhomogeneous Hamiltonians~\cite{PhysRevLett.123.130602}, is also in general a non-local and nonlinear expression in terms of $f$.
The effective velocity $v^{\mathrm{eff}}$ is defined by
\begin{equation}
  v^{\mathrm{eff}}
  := \frac{( \partial_\theta \epsilon)^{\mathrm{dr}} }{( \partial_\theta p)^{\mathrm{dr}} },
    \label{eq:effective_velocity}
\end{equation}
and the effective acceleration $a^{\mathrm{eff}}$ is defined by
\begin{equation}
  a^{\mathrm{eff}}
  := \frac{( -\partial_z \epsilon )^{\mathrm{dr}} }{( \partial_\theta p )^{\mathrm{dr}} },
    \label{eq:effective_acceleration}
\end{equation}
where $\epsilon = \epsilon(z, \theta)$ and $p = p(\theta)$ are the single-particle energy and momentum, respectively. 
Note that  $a^{eff}$ has the dimension of an acceleration if you divide by $\hbar m$ that is 1 in our scaling. Similar for $v^{eff}$. 
The exact expressions for $\epsilon$ and $p$ depend on the integrable model in question.
Meanwhile, the superscript $^{\mathrm{dr}}$ denotes the dressing operation, which describes the modification of single-particle properties following interactions with other particles. Given a state $f$, the dressing operation is defined as 
\begin{equation}
  g^{\mathrm{dr}} (\theta) =
  g(\theta) + \int_{\R} \mathrm{d}\theta' \: T(\theta - \theta') f(\theta')g^{\mathrm{dr}} (\theta').
  \label{eq:dressing}
\end{equation}
Here, $T(\theta)$ is the scattering kernel given by
\begin{equation}
  \label{def:kernel:T}
T(\theta) = \frac{\partial_\theta \varphi(\theta)}{2 \pi},
\end{equation}
with $\varphi (\theta)$ being the two-body scattering phase, which also depends on the specific model. 
Since the effective velocity (and in general the acceleration) are functionals of the distribution function $f$ itself, it turns out that the GHD advection equation \eqref{eq:GHD_equation} is a nonlinear kinetic equation with a self-consistent coupling.

Note, the dressing operation defines the function $g^{\mathrm{dr}}$ as the solution of an integral equation in the rapidity variable $\theta$, with the given source term $g$, and where the other variables, i.e. $(t,z)$, are fixed.
Introducing the following shorthand notation for the action of an integral operator $\widehat{K}$ with kernel $K$ on any function $g$,
\begin{equation}
  \big( \widehat{K} g \big) (\theta) \coloneqq \int_{\R} \mathrm{d} \theta^{\prime} \:
  K(\theta, \theta^{\prime}) g(\theta^{\prime}),
  \label{short_not_op}
\end{equation}
the dressing operation rewrites as
\begin{equation}
  g^{\mathrm{dr}} = g + \widehat{T}fg^{\mathrm{dr}},
  \label{eq:dressing_short}
\end{equation}
and formally the dressing operator is $(1-\widehat{T}f)^{-1}$, i.e.  $g^{\mathrm{dr}}=(1-\widehat{T}f)^{-1}g$, for any function $g$.\\ 

As for collisionless kinetic equations of Vlasov type, a conservative form of the GHD equations, written in terms of the quasi-particle density $\rho_\mathrm{p} =\rho_\mathrm{p} (t, z, \theta)$, exists and reads as
\begin{equation}
  \partial_{t} \rho_\mathrm{p}+ \partial_{z} \big( v^{\mathrm{eff}} \:
  \rho_\mathrm{p} \big) + \partial_{\theta} \big( a^{\mathrm{eff}} \: \rho_\mathrm{p} \big) =0.
  \label{eq:GHD_equation_conservative}
\end{equation}
Likewise, the effective velocity and acceleration can be expressed in terms of the quasi-particle density $\rho_{\mathrm{p}}$ yielding the equations
\begin{equation}
  v^{\mathrm{eff}}(t,z,\theta)=\partial_\theta \epsilon(z,\theta)+
  2 \pi\int_{\R} \mathrm{d} \theta^{\prime} \: T(\theta-\theta^{\prime})
  \rho_{\mathrm{p}}(t,z, \theta^{\prime})\big(v^{\mathrm{eff}}(t,z, \theta^{\prime})
  -v^{\mathrm{eff}}(t,z, \theta)\big),
  \label{eq:veff_rhop}
\end{equation}
and
\begin{equation}
  a^{\mathrm{eff}}(t,z, \theta)= -\partial_{z} \epsilon(z, \theta)+ 2 \pi\int_{\R} \mathrm{d}\theta^{\prime}
  \: T(\theta-\theta^{\prime}) \rho_{\mathrm{p}}
  (t,z, \theta^{\prime})\big(a^{\mathrm{eff}}(t,z, \theta^{\prime})-a^{\mathrm{eff}}(t,z, \theta)\big).
  \label{eq:aeff_rhop}
\end{equation}
In fact, the conservative form of the GHD equations \eqref{eq:GHD_equation_conservative}-\eqref{eq:aeff_rhop} is equivalent to the advective form of the GHD equations, which are constituted by equations \eqref{eq:GHD_equation}, \eqref{eq:effective_velocity} and \eqref{eq:effective_acceleration}.
Indeed, using the density of states $\rho_{\mathrm{s}}=\rho_{\mathrm{s}}(t,z,\theta)$, defined by
\begin{equation}
  \label{def:rhos}
  \rho_{\mathrm{s}} := \frac{1^{\mathrm{dr}}}{2\pi},
\end{equation}  
and the fundamental definition,
\begin{equation}
  \label{eqn:frhoprhos}
 f:= \frac{ \rho_{\mathrm{p}} }{ \rho_{\mathrm{s}}},
\end{equation}  
we obtain, from the definitions of the dressing operation \eqref{eq:dressing} and the scattering kernel $T$ given by \eqref{def:kernel:T}, that the equations \eqref{eq:effective_velocity}-\eqref{eq:effective_acceleration} for the advection field  $\mathbf F \equiv (v^{\mathrm{eff}},\, a^{\mathrm{eff}})^T$ is the same as \eqref{eq:veff_rhop}-\eqref{eq:aeff_rhop}.
To pass from the advective form \eqref{eq:GHD_equation} to the conservative form \eqref{eq:GHD_equation_conservative}, we use the relation \eqref{eqn:frhoprhos}, and the fact that the density of states $\rho_{\mathrm{s}}$, defined by \eqref{def:rhos}, satisfies the same conservation law as $\rho_{\mathrm{p}}$, namely,
\begin{equation}
  \partial_{t} \rho_\mathrm{s}+ \partial_{z} \big( v^{\mathrm{eff}} \,
  \rho_\mathrm{s} \big) + \partial_{\theta} \big( a^{\mathrm{eff}} \, \rho_\mathrm{s} \big) =0.
  \label{eq:CLrhos}
\end{equation}
In Appendix~\ref{Appendix:GHD} we show that the conservation law \eqref{eq:CLrhos} for  $\rho_{\mathrm{s}}$ can be recovered from the conservation law \eqref{eq:GHD_equation_conservative} for  $\rho_{\mathrm{p}}$.
In addition, in Appendix~\ref{Appendix:GHD} we show that the GHD equations preserve in time some integral invariants such as the total number of particles $N$, the total density of states $J$, the total energy $E$ and an infinite number of entropy $S$.
More precisely, for all $t\in\R_+$, we have,
\begin{eqnarray}
  && \frac{\mathrm{d}}{\mathrm{d}t}N(t)=0,  \quad \mbox{ with } \ \ \ N(t)= \int_\R \mathrm{d}z \int_\R \mathrm{d}\theta\,  \rho_\mathrm{p},\label{CL:N}\\
  && \frac{\mathrm{d}}{\mathrm{d}t}J(t)=0, \quad \ \mbox{ with } \ \ \ J(t)= \int_\R \mathrm{d}z \int_\R \mathrm{d}\theta\,  \rho_\mathrm{s},\label{CL:Gamma}\\
  && \frac{\mathrm{d}}{\mathrm{d}t}E(t)=0, \quad \ \mbox{ with } \ \ \  E(t)= \int_\R \mathrm{d}z \int_\R \mathrm{d}\theta\,
  \rho_\mathrm{p} \epsilon , \label{CL:E}\\
  && \frac{\mathrm{d}}{\mathrm{d}t}S(t)=0, \quad \ \mbox{ with } \ \ \ S(t)= \int_\R \mathrm{d}z \int_\R \mathrm{d}\theta\,  \rho_\mathrm{p}
  \Psi(f), \label{CL:S}
\end{eqnarray}  
where the function $\Psi: \R_+\mapsto \R$ is any smooth function.\\

The Lieb--Liniger model~\cite{LL1, LL2}, which describes a one-dimensional Bose gas with repulsive point-like interaction, is among the most common models treated with GHD.
Therefore, within the scope of this paper, we will concern ourselves only with this particular model.
However, by virtue of the universal formulation of GHD, all results and methods presented can be generalized in a straightforward manner to other integrable models~\cite{10.21468/SciPostPhysLectNotes.18}.
Given a gas of $N$ bosons, the Lieb--Liniger Hamiltonian reads, after a rescaling s.t. $\frac{\hbar^2}{2m} = 1$
\begin{equation}
    \mathcal { H } = - \sum _ { i } ^ { N } \frac { \partial ^ { 2 } } { \partial z _ { i } ^ { 2 } } + 2 c \sum _ { i < j } ^ { N } \delta \left( z _ { i } - z _ { j } \right),
    \label{eq:LiebLiniger_Hamiltonian}
\end{equation}
where $c>0$ is the coupling constant of the atoms and $z_i$ the position of the $i$'th boson.
For the Lieb--Liniger model, we have $\partial_\theta \varphi(\theta) = 2 c/(c^2 + \theta^2)$.
Further, the single-particle momentum and energy read
\begin{equation}
    p = p(\theta) := \theta ,
\end{equation}
and
\begin{equation}
  \label{def:hamiltonian}
\epsilon = \epsilon(z, \theta):= \theta^2 + W(z,\theta),
\end{equation}
respectively, where $W=W(z,\theta)$ is a general potential.
Most commonly the potential $W$ is an external trapping potential $W(z,\theta)=V(z)$ used to confine the atoms to a particular region in space.
In this case, the effective acceleration simplifies to $a^{\mathrm{eff}} (t,z,\theta) = a^{\mathrm{eff}} (z)=- \partial_z V(z)$~\cite{SciPostPhys.2.2.014}.
Note that for the Lieb-Liniger model we have $(\partial_\theta p)^{\mathrm{dr}} = 1^{\mathrm{dr}}$.

\subsection{The dissipative GHD equation }
\label{ss:ECD}
In the presence of diffusion or perturbations to integrability, the GHD evolution of the filling function $f$ takes the general form of a advective-dissipative equation
\begin{equation}
    \partial_{t} f+v^{\mathrm{eff}} \: \partial_{z} f+a^{\mathrm{eff}} \: \partial_{\theta} f = \mathcal{N}[f],
    \label{eq:GHD_equation_nonlin}
\end{equation}
where the source term $\mathcal{N}[f]$ is a nonlinear functional of the filling function $f$.
Unlike the GHD advection equation~\eqref{eq:GHD_equation}, the dissipative GHD equation~\eqref{eq:GHD_equation_nonlin}
no longer preserves the infinite set of conserved quantities, and thus enables thermalization. 
We will be studying two different nonlinear source terms $\mathcal{N}[f]$ in the following. First,
the diffusion operator $\mathcal{N} = \mathcal{D}$, which accounts for high-order corrections
to the GHD advection equation \cite{PhysRevLett.121.160603,NBD19,PhysRevLett.125.240604}, and second the Boltzmann-type collision integral $\mathcal{N} = \mathcal{I}$, which accounts for non-integrable perturbations to the system~\cite{Bastianello_2021, PhysRevB.101.180302, PhysRevLett.127.130601}.
Here, we will be treating an integrability-breaking perturbation in the form of 3D excitations of atoms in the Bose gas~\cite{10.21468/SciPostPhys.9.4.058, PhysRevLett.126.090602}.
Generally, in the presence of either of these nonlinear source terms,
the total number of particles $N$ and the total energy $E$ remain conserved, while entropies $S$ decrease
or increase monotonically, according to the choice of the sign convention (see \cite{NBD19, PhysRevLett.125.240604, PhysRevLett.126.090602}).
However, due to simplifications made in the construction of the collision integral for the 3D excitations, the total energy is not fully conserved for all interaction strengths of the Lieb--Liniger model.
Lastly, an important difference between these two dissipation operators
is that the diffusion operator acts in the variable $z$ locally (with a second-order derivative in this variable)
and non-locally in the variable $\theta$ (with some integral operators in this variable),
while the Boltzmann-type collision operator only acts in a non-local way in the variable $\theta$ (without any derivative in the variables $z$ and $\theta$).

\subsubsection{Diffusion operator}
\label{sss:Dop}

The GHD model can be extended by adding a diffusive term : While considering excitations with only one particle-hole pair above the reference state is sufficient to derive the GHD on the Euler scale, the diffusion arises due to simultaneous excitation of multiple (at least 2) particle-hole pairs. The resonance conditions (i.e., the momentum and energy conservation) for the scattering of two pairs are fulfilled in a general case, which leads to the violation of the genuinely ballistic propagation of quasiparticles. 
More detailed analysis \cite{NBD19} shows that the subleading to ballistic expansion order in this propagation is diffusive (its contribution to the two-point correlation functions of conserved charge grows linearly in time).

For the advective form of the extended GHD equation \eqref{eq:GHD_equation_nonlin},
the diffusion operator reads \cite{NBD19, PhysRevLett.125.240604},
\begin{equation}
  \mathcal{D}_f [f] = \frac{1}{2 \rho_{\mathrm{s}}}
  (1- f \widehat{T}) \: \partial_{z}\big((1-f \widehat{T})^{-1} \rho_{\mathrm{s}} \: \widehat{D} \: \partial_{z} f\big),
  \label{eq:diffusion_kernel}
\end{equation}
which, notably, depends on the $z$-derivative of the filling function.
Here, the kernel of the integral operator $\widehat{D}$ is given by
\begin{equation}
  D(\theta, \alpha)=
  \frac{1}{\rho_{\mathrm{s}}^{2}(\theta)} \Big( \delta(\theta-\alpha) \int_\R \mathrm{d}\gamma \: W(\gamma, \alpha) - W(\theta, \alpha) \Big),
  \label{def:kernel:D}
\end{equation}
where
\begin{equation}
  W(\theta, \alpha)=\rho_{\mathrm{p}}(\theta)\big(1-f(\theta)\big)
  \big[T^{\mathrm{dr}}(\theta, \alpha)\big]^{2}\big|v^{\mathrm{eff}}(\theta)-v^{\mathrm{eff}}(\alpha)\big|.
  \label{def:kernel:W}
\end{equation}
In \eqref{def:kernel:W} the dressed scattering kernel $T^{\mathrm{dr}}(\theta, \alpha)$ is the kernel
of the operator $(1-\widehat{T}f)^{-1}\widehat{T}$, i.e. $T^{\mathrm{dr}}(\theta, \alpha)=((1-\widehat{T}f)^{-1}T(\cdot - \alpha))(\theta)$,
or in other words the dressing of the function $T(\theta-\alpha)$ as a function of its first argument $\theta$.
Even if the variables $(t,z)$  are not explicitly written in  \eqref{def:kernel:D}-\eqref{def:kernel:W}, the kernels
$ D(\theta, \alpha)$ and  $W(\theta, \alpha)$ depend on the time variable $t$ and the position variable $z$, through the functions
$\rho_{\mathrm{s}}$, $\rho_{\mathrm{p}}$, and $f$.  In the presence of diffusion and an external potential $V(z)$,
the total number of quasi-particles $N$, defined  by \eqref{CL:N}, and the total energy $E$, defined by \eqref{CL:E}, 
remain conserved throughout evolution~\cite{PhysRevLett.125.240604} (see Appendix~\ref{appendix:IdGHDD} for a proof), while entropies $S$, defined
by \eqref{CL:S}, decrease monotonically \cite{NBD19}. For this, it is more convenient to work
with the conservative form of the GHD equations, namely
\begin{equation}
  \partial_{t} \rho_\mathrm{p}+ \partial_{z} \big( v^{\mathrm{eff}} \: \rho_\mathrm{p} \big)
  + \partial_{\theta} \big( a^{\mathrm{eff}} \: \rho_\mathrm{p} \big) = \mathcal{D}_{\rho_{\mathrm{p}}}[\rho_{\mathrm{p}}],
  \label{eqn:V-DCF}
\end{equation}
where the corresponding diffusion operator $\mathcal{D}_{\rho_{\mathrm{p}}}[\rho_{\mathrm{p}}]$ reads
\begin{equation}
  \mathcal{D}_{\rho_{\mathrm{p}}} [\rho_{\mathrm{p}}] = \frac{1}{2} \partial_{z}\big((1- f \widehat{T})^{-1}
  \rho_{\mathrm{s}}  \: \widehat{D} \:  \rho_{\mathrm{s}}^{-1} (1- f \widehat{T}) \: \partial_{z} \rho_{\mathrm{p}} \big).
   \label{eqn:DCF}
\end{equation}
Passing from the conservative form \eqref{eqn:V-DCF}-\eqref{eqn:DCF} to the advective form
\eqref{eq:GHD_equation_nonlin}, with $\mathcal{N}[f]=\mathcal{D}_f[f]$ and $ \mathcal{D}_f [f]$ given
by \eqref{eq:diffusion_kernel}, is detailed in \cite{NBD19}.

\subsubsection{Boltzmann-type collision integral}
\label{sss:Colop}
In an experimental setting, the Lieb-Liniger model can be realized by tightly confining an ultracold Bose gas along two of its axes.
Assuming the transverse confinement to be harmonic with trapping frequency $\omega_\perp$ and width $l_\perp = \sqrt{\hbar / m \omega_\perp}$,
the dynamics of the gas is effectively restricted to a single position axis if the transverse level spacing
$\hbar \omega_\perp$ exceeds all internal energy scales of the gas~\cite{PhysRevLett.87.130402, PhysRevLett.87.160405, doi:10.1126/science.1100700, PhysRevLett.105.265302}.
However, real systems are often only quasi one-dimensional, as excited states of the transverse confinement
can be populated through high energy collisions of atoms.
Such collisions break the integrability of the system and eventually lead to thermalization~\cite{10.21468/SciPostPhys.9.4.058}.
To account for non-integrable scattering processes within the framework of GHD, one can construct a Boltzmann-type collision integral~\cite{PhysRevB.101.180302, PhysRevLett.127.130601, Bastianello_2021}.
The construction of the collision integral accounting for transverse state-changing collisions is summarized in Appendix~\ref{appendix:collision}; the resulting collision integral is given by
\begin{equation} 
  \mathcal{I}[g]= \sum_{n = 1}^{2} \frac{1}{2} \big( \big\{ \mathcal{I}_{\mathrm{h}}^{+}[g]
    -\mathcal{I}_{\mathrm{p}}^{-}[g] \big\} + \big\{ \mathcal{I}_{\mathrm{h}}^{-}[g] -\mathcal{I}_{\mathrm{p}}^{+}[g]
    \big\} \nu_{n}^{\beta_n}  \big),
  \label{eq:collision_integral}
\end{equation}
where $g$ is respectively the filling function $f$ or the density of quasi-particles $\rho_{\mathrm{p}}$, depending
on whether the advective or conservative form of the extended GHD equations is considered~\cite{PhysRevLett.126.090602, PhysRevX.12.041032}.
Here, $\nu_n$ is the probability for an atom to be in the $n$'th transverse excited state, while $\beta_{1} = 2$
and $\beta_{2} = 1$ are the number of atoms changing state via the collisions. Note that the probability
 $\nu_n=\nu_n(t)$ is a function of the time variable $t$.
In addition, for solving the dissipative GHD equation with the Boltzmann-type collision integral~\eqref{eq:collision_integral},
one must also solve the rate equations for the excitation probabilities $\nu_{n} (t)$, namely 
\begin{equation} 
  \frac {\mathrm{d} \nu_n }{\mathrm{d} t}= \frac{1}{2} \beta_n \big\{ \Gamma_{\mathrm{h}}^+ -
    \Gamma_{\mathrm{p}}^+ \,\nu_{n}^{\beta_n} \big\}, 
    \label{eq:excitation_fraction} 
\end{equation}
where, for  $\alpha \in \{\mathrm{p},\, \mathrm{h}\}$, 
\begin{equation}
\Gamma_\alpha^+=(2N)^{-1}\int_\R \mathrm{d}z \int_\R \mathrm{d}\theta \, 
\rho_{\mathrm{s}} \,\mathcal{I}_\alpha^+[f] =
(2N)^{-1}\int_\R \mathrm{d}z \int_\R \mathrm{d}\theta \, 
\mathcal{I}_\alpha^+[\rho_{\mathrm{p}}].
\label{eq:exct_rate}
\end{equation}
The terms $\mathcal{I}_\alpha^\pm$ of eq.~\eqref{eq:collision_integral} are collision integrals of individual scattering processes, which for the advective form of the dissipative GHD equation \eqref{eq:GHD_equation_nonlin} can be written in the form 
\begin{equation} 
  \mathcal{I}_\alpha^\pm [f](\theta)= \frac{(2\pi)^2 \hbar}{m}\int_{\mathcal{R}_\pm } \mathrm{d}\theta^\prime \;
  \lvert\theta -\theta^\prime \rvert \: P_\updownarrow (\lvert\theta -\theta^\prime \rvert,\,
  \lvert\theta_\pm -\theta^\prime_\pm \rvert) \: f_\alpha (\theta )f_\alpha (\theta^\prime )
  f_{\bar \alpha}(\theta_\pm)f_{\bar \alpha }(\theta^\prime_\pm ) \:  \rho_{\mathrm{s}} (\theta^\prime)  \rho_{\mathrm{s}} (\theta_\pm)  \rho_{\mathrm{s}} (\theta_\pm^\prime).
  \label{eq:collision_channel}
\end{equation}
Here, $\bar \alpha = {\mathrm{h}}$, for $\alpha = {\mathrm{p}}$ and vice versa, while $f_{\mathrm{p}} = f$ and $f_{\mathrm{h}} = 1 - f$.
Further, $P_\updownarrow (\theta_1,\, \theta_2)=4 c^2\theta_1\theta_2 /[\theta_1^2\theta_2^2 +c^2(\theta_1+\theta_2 )^2]$
is the scattering probability, while
$\theta_\pm = \frac 12 (\theta +\theta^\prime )+\frac 12 (\theta -\theta^\prime )\sqrt{1\pm 8/[(\theta -\theta^\prime )l_\perp ]^2}$ 
and
$\theta_\pm^\prime = \frac 12 (\theta +\theta^\prime )-\frac 12 (\theta -\theta^\prime )\sqrt{1\pm 8/[(\theta -\theta^\prime )l_\perp ]^2}$
are the rapidities after a collision leading to excitation
or de-excitation of the transverse states, respectively.
The integration ranges in equation~\eqref{eq:collision_channel} are the following: 
$\mathcal{R}_+$ is the whole real axis, and $\mathcal{R}_-$ is comprised of those real values of $\theta^\prime $,
which yield real $\theta_- $ and $\theta_-^\prime $, i.e. 
$\mathcal{R}_- =\{ \theta^\prime : \theta^\prime < \theta -\sqrt 8/l_\perp \}\cup \{ \theta^\prime : \theta^\prime > \theta +\sqrt 8/l_\perp \}$.
Even if the variables $(t,z)$ are not explicitly written in eqs.~\eqref{eq:collision_integral}-\eqref{eq:collision_channel}, the collision integrals $\mathcal{I}_\alpha^\pm [f](\theta)$ depend on the time variable $t$ and position variable $z$ through the functions $\rho_{\mathrm{s}}$ and $f$.

Boltzmann-type collision integrals are explicitly constructed to preserve the total number of particles $N$.
Generally, this is also the case for the total energy $E$, however, in the derivation of the collision integral~\eqref{eq:collision_channel} the backflow, describing a shift in all local rapidities following an excitation~\cite{LL2}, was neglected.
Hence, the associated energy shift is not accounted for and the total energy is not conserved, although for most experimentally realizable parameters of the Lieb--Liniger model the error is not too severe~\cite{PhysRevLett.126.090602}.
For more details, see Appendix~\ref{appendix:collision}.

Again, the conservation of particle number is most obvious when examining its equivalent conservative form given by 
\begin{equation}
  \partial_{t} \rho_\mathrm{p}+ \partial_{z} \big( v^{\mathrm{eff}}
  \: \rho_\mathrm{p} \big) + \partial_{\theta} \big( a^{\mathrm{eff}} \: \rho_\mathrm{p} \big)
  = \mathcal{I}[\rho_{\mathrm{p}}].
  \label{eqn:V-BCF}
\end{equation}
The form of the collision integral $\mathcal{I}[\rho_{\mathrm{p}}]$, which is originally constructed in \cite{PhysRevLett.126.090602},
has the same form as equation~\eqref{eq:collision_integral}, however,
the collision integrals of the individual collision channels now read
\begin{equation} 
  \mathcal{I}_\alpha ^\pm [\rho_{\mathrm{p}}](\theta) = \frac{(2\pi )^2\hbar}{m}\int_{\mathcal{R}_\pm }
  \mathrm{d}\theta^\prime \, |\theta -\theta^\prime |P_\updownarrow (|\theta -\theta^\prime |,\,
  |\theta_\pm -\theta^\prime_\pm |) \rho_\alpha (\theta )\rho_\alpha (\theta^\prime )
  \rho_{\bar \alpha}(\theta _\pm )\rho_{\bar \alpha }(\theta^\prime_\pm ),
    \label{eq:collision_channel_rho}
\end{equation} 
where $\bar \alpha  = \mathrm{h}$, for $\alpha = \mathrm{p}$ and vice versa, and with  $ \rho_{\bar \alpha} = \rho_{\mathrm{s}} - \rho_\alpha$.

\section{Analytic time derivatives of the advection fields} \label{s:ATDAF}
In this section, we derive analytic time derivatives of the advection fields $v^{\mathrm{eff}}$ and $a^{\mathrm{eff}}$,
which will be used to high-order schemes in time for solving the GHD advection equation~\eqref{eq:GHD_equation}. 
Indeed, knowing the advection fields and its time derivatives up to high order, we will obtain a high-order
approximation of the advection fields  at any time by using a time-Taylor series expansion (see also \cite{SP92} where some Taylor series expansions are used).
For this, we define auxillary advection fields  $w^{\mathrm{eff}}$  and  $b^{\mathrm{eff}}$ as follows,
\begin{equation*}
  w^{\mathrm{eff}} \coloneqq 1^{\mathrm{dr}} v^{\mathrm{eff}}, \quad \mbox{ and } \quad
  b^{\mathrm{eff}} \coloneqq 1^{\mathrm{dr}} a^{\mathrm{eff}}.
  \label{eqn:AuxAV}
\end{equation*}
Given the definition of the dressing equation~\eqref{eq:dressing}, $w^{\mathrm{eff}}$ and $b^{\mathrm{eff}}$ solve the following equations,
\begin{align*}
\begin{split}
w^{\mathrm{eff}} (\theta) = \partial_\theta \epsilon + \int_\R \mathrm{d}\theta' \: T(\theta - \theta') f(\theta') w^{\mathrm{eff}} (\theta'),
\\
b^{\mathrm{eff}} (\theta) = - \partial_z \epsilon  + \int_\R \mathrm{d}\theta' \: T(\theta - \theta') f(\theta') b^{\mathrm{eff}} (\theta'),
\end{split}
\end{align*}
where $\epsilon = \epsilon(\theta, z)= \theta^2 + V(z)$ is the one-particle Hamiltonian.
Note that the filling function $f$ is also a function of time and position, i.e.  $f = f(t, z, \theta)$;
although these arguments will be omitted in the following whenever they appear the same on either side of an equality. 
Differentiating $w^{\mathrm{eff}}$ in time, we obtain
\begin{equation*}
  \partial_t w^{\mathrm{eff}} (\theta) = \int_\R \mathrm{d}\theta' \: T(\theta - \theta') \partial_t f(\theta')
  w^{\mathrm{eff}} (\theta') + \int_\R \mathrm{d}\theta' \: T(\theta - \theta') f(\theta') \partial_t w^{\mathrm{eff}} (\theta').
\end{equation*}
Here, $\partial_t f(\theta')$ is given through the GHD advection equation~\eqref{eq:GHD_equation}, implying
that the time derivative of $f$ is expressed in terms of the  position and rapidity derivatives of $f$, and of 
other known quantities, such as the advection fields $(v^{\mathrm{eff}}, \,a^{\mathrm{eff}})$.
Defining the source term $S_w$ as
\begin{equation*}
    S_w (\theta) \coloneqq \int_\R \mathrm{d}\theta' \: T(\theta - \theta') \partial_t f(\theta') w^{\mathrm{eff}} (\theta'),
\end{equation*}
we see that
\begin{equation*}
   \partial_t w^{\mathrm{eff}} (\theta) = S_w^{\mathrm{dr}} (\theta).
\end{equation*}
Likewise, a source term for $b^{\mathrm{eff}}$ can be defined such that
\begin{equation*}
    \partial_t b^{\mathrm{eff}} (\theta) = S_b^{\mathrm{dr}} (\theta),
\end{equation*}
where
\begin{equation*}
    S_b (\theta) = \int_\R \mathrm{d}\theta' \: T(\theta - \theta') \partial_t f(\theta') b^{\mathrm{eff}} (\theta').
\end{equation*}
Knowing the derivatives $\partial_t w^{\mathrm{eff}}$ and $\partial_t b^{\mathrm{eff}}$, the time derivatives of the advection fields can be computed following
\begin{align}
\begin{split}
\partial_t v^{\mathrm{eff}}  = \frac{\partial_t w^{\mathrm{eff}}  - \partial_t 1^{\mathrm{dr}}  v^{\mathrm{eff}} }{ 1^{\mathrm{dr}} },
\\
\partial_t a^{\mathrm{eff}}  = \frac{\partial_t b^{\mathrm{eff}}  - \partial_t 1^{\mathrm{dr}}  a^{\mathrm{eff}} }{ 1^{\mathrm{dr}} },
\end{split}
\label{eq:timedwb}
\end{align}
where the unit-cell $1^{\mathrm{dr}}$ obeys the conservation law
\begin{equation}
  \partial_{t}  1^{\mathrm{dr}} + \partial_{z} \big( v^{\mathrm{eff}} \:
  1^{\mathrm{dr}} \big) + \partial_{\theta} \big( a^{\mathrm{eff}} \: 1^{\mathrm{dr}} \big) = 0.
\label{eq:clrs}
\end{equation}
Thereby, knowing both $1^{\mathrm{dr}}$ and the advection fields $(v^{\mathrm{eff}}, \,a^{\mathrm{eff}})$,
we can compute $\partial_t 1^{\mathrm{dr}}$ from \eqref{eq:clrs} and in turn the time derivatives of the advection fields
$(v^{\mathrm{eff}}, \,a^{\mathrm{eff}})$ from \eqref{eq:timedwb}.
Higher-order time derivatives of the advection fields can be derived in similar fashion.
If we let $S_w^{(1)} \equiv S_w$, we can construct higher-order source terms fulfilling
\begin{equation*}
  \partial_t^{(k)} w^{\mathrm{eff}} (\theta) = \big( S_w^{(k)} \big)^{\mathrm{dr}} (\theta),
\end{equation*}
where $ S_w^{(k)} (\theta)$ depends on all previous source terms $ S_w^{(j)} (\theta)$ with $j < k$. Of course, we have similar
expressions for  $\partial_t^{(k)} b^{\mathrm{eff}}(\theta)$ in terms of $ S_b^{(k)} (\theta)$.  
For example, the second-order source term $S_w^{(2)} (\theta) $ reads
\begin{equation}
  S_w^{(2)} (\theta) = \int_\R \mathrm{d}\theta' \: T(\theta - \theta') \partial_t^{(2)} f(\theta') w^{\mathrm{eff}} (\theta')
  + 2 \int_\R \mathrm{d}\theta' \: T(\theta - \theta') \partial_t f(\theta') \partial_t w^{\mathrm{eff}} (\theta').
  \label{eqn:sto2}
\end{equation}
In order to obtain high-order time derivatives of $f$, such as $\partial_t^{(2)} f$ appearing in \eqref{eqn:sto2},
we use the successive time derivatives of the GHD advection equation~\eqref{eq:GHD_equation}, where for any mixed
derivative of $f$, containing some time derivatives, each time derivative of $f$ is replaced by the  $(z,\theta)$-derivative of $f$ through
the GHD advection equation~\eqref{eq:GHD_equation}. In such a way, high-order time derivatives of $f$ can be expressed
only in terms  of the $(z,\theta)$-derivatives of $f$ and  of other known quantities, such as the previously calculated
time derivatives of the advection fields $(v^{\mathrm{eff}}, \,a^{\mathrm{eff}})$.
In the same spirit, we use the successive time derivatives of
equations \eqref{eq:timedwb}-\eqref{eq:clrs} to obtain the high-order time derivatives of  $(v^{\mathrm{eff}}, \,a^{\mathrm{eff}})$ from those of
$(w^{\mathrm{eff}}, \,b^{\mathrm{eff}})$.
Finally, knowing the advection fields $(v^{\mathrm{eff}}, \,a^{\mathrm{eff}})$ at time $\tau$ and their time derivatives up to the order $k$,
we can obtain a high-order approximation of the advection fields at any time $t$ following the time-Taylor series expansion,
\begin{align}
\label{eq:velocity_expansion}
\begin{split}
  v^{\mathrm{eff}} (t) &= v^{\mathrm{eff}} (\tau) + (t - \tau) \:\partial_t v^{\mathrm{eff}} (\tau)
  + \ldots + \frac{(t - \tau)^k}{k!} \:\partial_t^{(k)} v^{\mathrm{eff}} (\tau),
\\
a^{\mathrm{eff}} (t) &= a^{\mathrm{eff}} (\tau) + (t - \tau) \:\partial_t a^{\mathrm{eff}} (\tau)
+ \ldots + \frac{(t - \tau)^k}{k!} \:\partial_t^{(k)} a^{\mathrm{eff}} (\tau).
\end{split}
\end{align}

\section{Solving the dressing equation} \label{s:dressing}
The dressing operation, given by equation~\eqref{eq:dressing}, is ubiquitous in GHD,
as it describes how a single particle, or \textit{bare} function is modified in the presence of collective interactions.
Importantly, a dressed quantity becomes a function of time and position (and of course rapidity) through the dependence on the filling function $f(t,z,\theta)$.
Therefore, at every step of time propagation, the dressing operation presents a new set of equations to be solved.

To solve the dressing equation, we first discretize position and rapidity creating the grids
$z_i \in [z_1, z_2, \ldots, z_{N_z}]$ and $\theta_j \in [\theta_1, \theta_2, \ldots, \theta_{N_\theta}]$.
Although the rapidity integral in equation~\eqref{eq:dressing} is taken over the whole real axis,
in practice it is sufficient to integrate over the interval $[\theta_\mathrm{min} , \theta_\mathrm{max}]$ containing
the support in $\theta$ of the filling function $f$.
Since the dressing operation is completely local in position,
the equation can be solved independently for every point in position $z_i$.
Then, in discretized form, the dressing equation for a single point in position reads
\begin{equation}
  g^{\mathrm{dr}} (\theta_i) = g(\theta_i) + \sum_{j = 1}^{N_\theta} \mathrm{w}_j \, T(\theta_i - \theta_j) \, f(\theta_j) g^{\mathrm{dr}} (\theta_j), 
  \label{eq:dressing_discretized}
\end{equation}
where $\mathrm{w}_j$ are the quadrature weights for the corresponding discretization scheme of the rapidity axis. 
Note that the position and time arguments have been omitted entirely for a more compact notation, as they are the same on either side of the equality sign. 
The discretized form of the dressing equation~\eqref{eq:dressing_discretized} can be rewritten as
\begin{equation}
    g(\theta_i) = \sum_{j = 1}^{N_\theta} \big[ \delta_{ij} -  \mathrm{w}_j \, T(\theta_i - \theta_j) \, f(\theta_j) \big] g^{\mathrm{dr}} (\theta_j), 
    \label{eq:dressing_linear}
\end{equation}
where $\delta_{ij}$ is the standard Kronecker symbol.
Hence, equation~\eqref{eq:dressing_linear} can be written in compact form as the matrix product $g = \mathbf U \cdot  g^{\mathrm{dr}}$,
where the elements of the matrix $\mathbf U$ are $\mathbf U_{ij} = \delta_{ij} -  \mathrm{w}_j \, T(\theta_i - \theta_j) \, f(\theta_j) $.
Applying the dressing operation numerically thus amounts to solving the set of linear equations of
equation~\eqref{eq:dressing_linear} either through inverting the matrix $\mathbf U$ or,
as in the case of the iFluid framework~\cite{10.21468/SciPostPhys.8.3.041},
using a subroutine such as MATLAB \texttt{mldivide} function~\cite{MATLAB:2020}, capable of leveraging symmetries of the matrix.

\begin{figure}[t]
    \centering
    \includegraphics{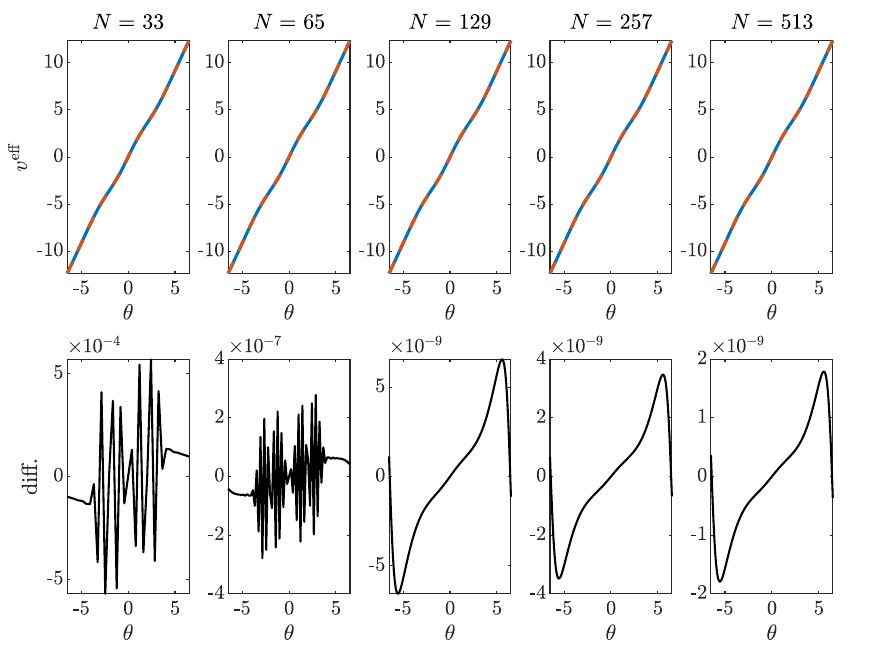}
    \caption{\label{fig:rapidity_quadrature}
      Comparison of effective velocity~\eqref{eq:effective_velocity} calculated for the test state~\eqref{eq:quadrature_test_state} using
      the trapezoidal (blue) and Simpson's (red) rule to discretize the integral over rapidities featuring
      in the dressing equation~\eqref{eq:dressing}. The effective velocity is calculated and plotted for increasingly
      fine discretization of the rapidity axis ($N_\theta$ is the number of grid points). Below, the difference between the two results are plotted.
    }
\end{figure}

In general, when solving the dressing equation it is most convenient to employ the same discretization scheme used in the backward semi-Lagrangian method.
Otherwise interpolation of the filling function $f$ between the two grids must be performed at every evaluation of the dressing equation.
Hence, for the entirety of this work, a uniform discretization of both the position and rapidity axis will be employed.
For such discretization several quadrature rules exist to approximate this integral.
Here, we test two of the most standard: the trapezoidal rule and the Simpson's rule.
Following the trapezoidal rule, the quadrature weights $\mathrm{w}_j$ read
\begin{equation*}
  \{ \mathrm{w}_1, \mathrm{w}_2, \mathrm{w}_3, \ldots,  \mathrm{w}_{N_\theta-1}, \mathrm{w}_{N_\theta} \}
  = \frac{(\theta_\mathrm{max} - \theta_\mathrm{min})}{2 (N_\theta-1)} \{ 1, 2, 2, \ldots, 2, 1 \},
\end{equation*}
while for the Simpson's rule they read
\begin{equation*}
\{ \mathrm{w}_1, \mathrm{w}_2, \mathrm{w}_3, \mathrm{w}_4, \mathrm{w}_5, \ldots, \mathrm{w}_{N_\theta-2}, \mathrm{w}_{N_\theta-1}, \mathrm{w}_{N_\theta} \}
  = \frac{(\theta_\mathrm{max} - \theta_\mathrm{min})}{3 (N_\theta-1)} \{ 1, 4, 2, 4, 2 \ldots, 2, 4, 1 \}.
\end{equation*}
To compare the two quadrature rules, we compute the effective velocity~\eqref{eq:effective_velocity}
at different grid resolutions of the reference state
\begin{equation}
  f_0 (\theta) = 0.9 \left[ \exp \left( - \frac{\left( \theta - \theta_\mathrm{B} \right)^2 }{2 \sigma} \right)
    + \exp  \left( - \frac{\left( \theta + \theta_\mathrm{B} \right)^2 }{2 \sigma} \right) \right], 
    \label{eq:quadrature_test_state}
\end{equation}
where $\theta_\mathrm{B} = 2$ and $\sigma = 1 / \sqrt{2}$.
The results can be seen in Fig.~\ref{fig:rapidity_quadrature}.
Although a clear systematic difference between the two quadrature rules is apparent, the difference is nonetheless
almost vanishing for $N_\theta > 100$. Then, in order to keeps matters simple,
for the entirety of this work the trapezoidal rule will be employed.

\section{The backward semi-Lagrangian (BSL) method} \label{s:BSL}
The characteristic equations associated with the first-order PDE~\eqref{eq:GHD_equation} are
\begin{subequations}
\begin{align}
     \frac{\mathrm{d}Z(t)}{\mathrm{d}t} = v^{\mathrm{eff}}(t, Z(t), \Theta(t)) \quad , \quad Z(t; t, z, \theta) = z,\\
     \frac{\mathrm{d}\Theta(t)}{\mathrm{d}t} = a^{\mathrm{eff}}(t, Z(t), \Theta(t)) \quad , \quad \Theta(t; t, z, \theta) = \theta,
\end{align}
\label{eqn:EqChar}
\end{subequations}
while the filling function $f$ is constant along the characteristics, i.e.,
\begin{equation}
    f(t, z, \theta)=f(s, Z(s; t, z, \theta), \Theta (s;t, z, \theta)), \quad \forall s\in \R.
    \label{eq:GHD_implicit_solution}
\end{equation}
We recall that the advection field $\mathbf F := (v^{\mathrm{eff}},\, a^{\mathrm{eff}})^T$ is
a functional of the filling function $f$,
through the integral equations \eqref{eq:effective_velocity}-\eqref{eq:effective_acceleration}. 
Using the Cauchy--Lipschitz--Picard--Lindel\"of theorem, existence and uniqueness
of solutions $(Z(t),\Theta(t))$ for ODEs \eqref{eqn:EqChar} hold true provided that the advection field  $\mathbf F (t,z,\theta)$ is bounded in the time variable $t$ and Lipschitz continuous in the phase-space variables $(z,\theta)$. Here, such a regularity is assumed. Nevertheless, as $\mathbf F$ is defined self-consistently through a nonlinear functional of $f$, to determine the regularity properties of $\mathbf F$, we must study, in a convenient functional framework, the Cauchy problem for the GHD equations, which is out of the scope of this paper.

\begin{figure}
    \centering
    \includegraphics{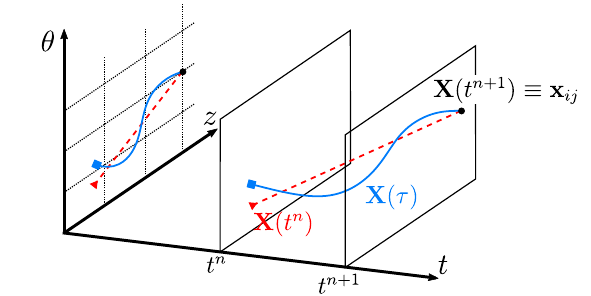}
    \caption{\label{fig:BSL_illustration}
    Illustration of a single propagation step using a backward semi-Lagrangian scheme.
    The planes of the phase-space $(z,\theta)$ at times $t^n$ and $t^{n+1}$ are drawn as rectangles, while their projection including a mesh indicating the discretization of the phase-space is shown in the back.
    Following the Lagrangian formalism, distribution $f$, whose values are known on the mesh at the starting time $t^n$, is constant along the characteristic $\mathbf X (\tau)$ for $\tau \in [ t^n, t^{n+1} ]$ (solid blue line).      
    In BSL schemes, the arrival point of the characteristic at the next time step $\mathbf X (t^{n+1})$ is assumed to coincide with the mesh $\mathbf x_{ij} = (z_i, \theta_j)$, while the departure point $\textbf X(t^{n})$ (marked as a square) is unknown.
    The solution of the corresponding ODEs~\eqref{eq:BSL_ODE} can be approximated using various numerical schemes (dashed red line) to obtain an estimate of the departure point (triangle).
    }
\end{figure}

Several methods for numerically solving equation~\eqref{eq:GHD_equation} exist.
In purely Lagrangian methods, the implicit solution~\eqref{eq:GHD_implicit_solution}
is utilized by considering the initial state $f(0, z, \theta)=f_0(z,\theta)$
as a distribution of fluid packages, whose trajectories, following time evolution, are encoded in the characteristics defined by equations \eqref{eqn:EqChar}.
However, following dynamics, the fluid packages will be scattered around the entire phase space,
which may result in the mesh needed for evaluating the effective velocity becoming heavily distorted and irregular~\cite{XIU2001658}.
Backward Semi-Lagrangian (BSL) schemes~\cite{BesseSonnenJCP03} 
circumvent this issue by assuming the \textit{arrival point} of the characteristics at each time step
to coincide with a pre-defined grid, effectively re-meshing at every time step~\cite{Staniforth1991}.
The BSL schemes then proceed in the following way. First, we discretize position and rapidity,
creating the grids $z_i \in \{ z_1, z_2, \ldots, z_{N_z} \}$ and $\theta_j \in \{ \theta_1, \theta_2, \ldots, \theta_{N_\theta} \}$.
For a single time step $\Delta t$, the GHD advection equation~\eqref{eq:GHD_equation}, integrated in time, writes as
\begin{equation}
  \label{eqn:fcac}
  f(t^{n+1}, z_i, \theta_j) = f(t^n, Z(t^n; t^{n+1}, z_i, \theta_j), \Theta(t^n; t^{n+1}, z_i, \theta_j)),
\end{equation}
where $t^n = n \Delta t$. 
To compute $f(t^{n+1})$ at the grid points $(z_i, \theta_j)$, we consider the characteristics which arrive at time $t^{n+1}$
on the grid points $(z_i, \theta_j)$. Then, the  position and rapidity couple $\big( Z( t^n;t^{n+1}, z_i, \theta_j), \Theta( t^n; t^{n+1}, z_i, \theta_j) \big)$ denotes the \textit{departure point}
of the fluid package at time $t^n$, which at the next time  $t^{n+1}$ will arrive at  the grid point $(z_i, \theta_j)$.
For a more compact notation we write the phase-space coordinate as $\mathbf x \equiv (z, \theta)^T$, the characteristics as
$\mathbf X (t^n) \equiv \big(Z(t^n; t^{n+1}), \Theta(t^n;t^{n+1}) \big)^T$, and the advection fields as
$\mathbf F \equiv (v^{\mathrm{eff}},\, a^{\mathrm{eff}})^T$. Hence, a solution to equation~\eqref{eq:GHD_equation} is obtained from \eqref{eqn:fcac}
and by following the characteristics $\textbf X$ backward in time, starting from the grid points $\mathbf x_{ij}$, i.e. by solving,
\begin{equation}
    \begin{dcases}
        \frac{\mathrm{d} \mathbf X (t)}{\mathrm{d} t} &  = \mathbf F(t, \mathrm{X}(t)) \\
        \mathbf X (t^{n+1}) & \equiv \mathbf X^{n+1} \coloneqq \mathbf x_{ij}.
   \end{dcases}
   \label{eq:BSL_ODE}
\end{equation}
Note that the characteristics in general do not coincide with the grid points.
So we must employ interpolation in order to obtain values of the filling function $f$ or the advection field $\textbf F$
at any point in the phase space. With this in mind, a single propagation step following a BSL scheme can be divided into two sub-steps:
First, the origin of the characteristics $\mathbf X (t^n)$ is computed. An illustration of this can be seen in Fig.~\ref{fig:BSL_illustration}.
Next, the filling function at the origin $f(t^n, \textbf X(t^n))$ is reconstructed using interpolation
from the known values on the mesh. Since the value of the filling function is constant along the characteristics,
the state at time $t^{n+1}$ is obtained as follows,
\begin{equation*}
    f^{n+1}(\mathbf{x}_{ij})= \Pi_{h} f^{n}(\mathbf{X}^{n}),
\end{equation*}
where $\Pi_{h}$ denotes the interpolation operator.
The choice of $\Pi_{h}$ is crucial for the accuracy and stability of the scheme \cite{Bes04, Bes08, BM08}.
In the following, we will employ an interpolation scheme based on a cubic spline using not-a-knot end conditions \cite{DeB01, Sch07}.

\section{High-order schemes in time to solve the BSL method} \label{s:schemes}

In this section we present a number of high-order schemes for solving the advective and dissipative GHD equations.
First, in Section~\ref{ss:IERKSG},  we focus on the purely advective case, for which we develop a set of novel implicit/explicit Runge--Kutta semi-Lagrangian methods.
These methods employ the exact time-derivatives of the GHD advection fields derived in Section~\ref{s:ATDAF} to estimate the advection fields at intermediate time steps.
For comparison, we introduce both Adams--Moulton schemes in Section~\ref{ss:AMSLS} and leap-frog schemes in Section~\ref{ss:LF}. Finally, in Section~\ref{ss:BSLH} we present a number of BSL-hybrid schemes for solving the dissipative GHD equation.

\subsection{Implicit/explicit Runge--Kutta semi-Lagrangian schemes}
\label{ss:IERKSG}
Runge--Kutta methods are a family of methods that employ a number of intermediate time steps in order to obtain a high-order
approximation for the solution of ordinary differential equations, such as the one of equation~\eqref{eq:BSL_ODE}.
Suppose that one knows the advection field $\mathbf F(t, \cdot)$ at any time $t$. 
For the purpose of solving the GHD advection equation~\eqref{eq:GHD_equation}, and following the recursive method
of Section~\ref{s:ATDAF}, the advection field $\mathbf F(t, \cdot)$
can be predicted by computing successively the time derivatives $\partial_t^{(k)} \mathbf F(t^n, \cdot)$, for $k \geq 0$,
and then using the time-Taylor series expansion~\eqref{eq:velocity_expansion} to reconstruct $\mathbf F(t, \cdot)$
at any time $t \in [t^n, t^{n+1}]$ (see Section~\ref{s:ATDAF} for more details).
Then, using the Runge--Kutta schemes of order $s$, that we formally write as
$\mathrm{RK}_{\mathbf F}^{(s)} (\Delta t; t^{n}, t^{n+c_2} , t^{n+c_3}, \ldots , t^{n+c_s}  ; \mathbf X (t^{n}))$,
with $0 < c_i < 1$ and $i \in \{ 2, \ldots, s \}$, we obtain
\begin{equation}
\begin{aligned}
    \mathbf X (t^{n+1})- \mathbf X (t^{n}) &=\int_{t^{n}}^{t^{n+1}} \mathbf F(s , \mathbf X (s))\, \mathrm{d}s \\
    & \simeq \Delta t \: \mathrm{RK}_{\mathbf F}^{(s)} \big(\Delta t; t^{n}, t^{n+c_2} , t^{n+c_3}, \ldots , t^{n+c_s}  ; \mathbf X (t^{n})\big).
\end{aligned}
\label{eq:RK_scheme}
\end{equation}
The function $\mathrm{RK}_{\mathbf F}^{(s)} (\Delta t; t^{n}, t^{n+c_2} , t^{n+c_3}, \ldots , t^{n+c_s}  ; \cdot)$
is known explicitly in terms of the functions $\mathbf F(t^n, \cdot)$, $\mathbf F(t^{n + c_2}, \cdot)$, $\ldots$, $\mathbf F(t^{n + c_s}, \cdot)$.
Then, to obtain the departure points $\mathbf X (t^{n})$ from equation~\eqref{eq:RK_scheme},
one must solve the following fixed-point problem corresponding to finding the zeroes of the following nonlinear function,
\begin{equation}
  \mathbf X \to G(\mathbf X) \coloneqq \mathbf x_{ij} - \mathbf X - \Delta t \:
  \mathrm{RK}_{\mathbf F}^{(s)} \big(\Delta t; t^{n}, t^{n+c_2} , t^{n+c_3}, \ldots , t^{n+c_s}  ; \mathbf X \big).
    \label{eq:RK_fixedpoint}
\end{equation}
The problem, $G(\mathbf X) = 0$, can be solved by using the Picard iteration (first order) or the  Newton iteration (second order).
Here, we employ the former, which usually converges in less than five iterations.
While the fixed-point problem of equation~\eqref{eq:RK_fixedpoint} is \textit{implicit}, one can also construct
\textit{explicit} Runge--Kutta schemes such as
\begin{equation*}
    \mathbf X (t^n) \simeq \mathbf x_{ij} - \Delta t \: \mathrm{RK}_{\mathbf F}^{(s)} \big(\Delta t; t^{n}, t^{n+c_2} , t^{n+c_3}, \ldots , t^{n+c_s}  ; \mathbf x_{ij} \big).
\end{equation*}
Note that the functions $\mathrm{RK}_{\mathbf F}^{(s)}$ used in the implicit and explicit schemes are usually not the same.
Explicit schemes rely only on known quantities and do not require to solve any fixed-point problem by successive iterations,
whereby they often are faster than implicit schemes. However, given two schemes at the same order, the implicit scheme is often more accurate,
allowing for overall larger time steps.

We end this section by giving a convergence criteria for the Picard iteration used to solve the implicit scheme \eqref{eq:RK_fixedpoint}, by following the method described in \cite{SP92}.
Let $\widetilde{\mathbf F}$ be an approximation of $\mathbf F$, which is obtained from the phase-space discretization of $f$, the numerical approximation of the dressing operation (see Section~\ref{s:dressing}) and the time-Taylor series expansion of the advection fields \eqref{eq:velocity_expansion} (see Section~\ref{s:ATDAF}). Then, using the Runge--Kutta scheme of order $\mathcal{O}(\Delta t^{s+1})$, given by \eqref{eq:RK_scheme}, we obtain
\begin{equation}
\label{eqn:RKs}
\mathbf X + \mathcal{O}\big(\Delta t^{s+1}, \, 
\Delta t \big\| \mathrm{RK}_{\mathbf F}^{(s)}-  \mathrm{RK}_{\widetilde{\mathbf F}}^{(s)}\big\|\big)
= \mathbf x_{ij}- \Delta t \:
\mathrm{RK}_{\widetilde{\mathbf F}}^{(s)} \big(\Delta t; t^{n}, t^{n+c_2} , t^{n+c_3}, \ldots , t^{n+c_s}  ;\mathbf  X \big).
\end{equation}
The Picard iteration can be compactly written as
\begin{equation}
\label{eqn:RKsPicard}
\mathbf X^\nu + \mathcal{O}\big(\Delta t^{s+1}, \, 
\Delta t \:\big\| \mathrm{RK}_{\mathbf F}^{(s)}-  \mathrm{RK}_{\widetilde{\mathbf F}}^{(s)}\big\|\big)
= \mathbf x_{ij}- \Delta t \:
\mathrm{RK}_{\widetilde{\mathbf F}}^{(s)} \big(\Delta t; t^{n}, t^{n+c_2} , t^{n+c_3}, \ldots , t^{n+c_s}  ;\mathbf  X^{\nu-1} \big),
\end{equation}
where the index $\nu$ numbers successive iterations.
Subtracting \eqref{eqn:RKsPicard}  from \eqref{eqn:RKs},
taking the norm of the resulting difference, and assuming
the differentiability of the function 
$\mathbf X\mapsto \mathrm{RK}_{\widetilde{\mathbf F}}^{(s)}(\ldots; \mathbf X)$, we obtain
\begin{eqnarray*}
\Big \| \mathbf X-\mathbf X^\nu + \mathcal{O}\big(\Delta t^{s+1}, \, 
\Delta t \:\big\| \mathrm{RK}_{\mathbf F}^{(s)}-  \mathrm{RK}_{\widetilde{\mathbf F}}^{(s)}\big\|\big)\Big \|
&=&\Delta t\: \big \|
\mathrm{RK}_{\widetilde{\mathbf F}}^{(s)}(\ldots;\mathbf X)
-\mathrm{RK}_{\widetilde{\mathbf F}}^{(s)}(\ldots,\mathbf X^{\nu-1})
\big\| \\ \nonumber
&\leq& \Delta t \: \bigg\| \frac{\partial \mathrm{RK}_{\widetilde{\mathbf F}}^{(s)}}{ \partial \mathbf X}\bigg\| \|\mathbf X-\mathbf X^{\nu-1}  \|.
\end{eqnarray*}
This last inequality implies that the Picard iteration converges 
if the following condition
\begin{equation*}
\Delta t \: \bigg\| \frac{\partial \mathrm{RK}_{\widetilde{\mathbf F}}^{(s)}}{ \partial \mathbf X}\bigg\| <1,
\end{equation*}
is satisfied.

\subsubsection{First-order schemes (RK1)}
Here, we consider a first order scheme, just to demonstrate how to compute the origin of the characteristics.
For most practical purposes first-order schemes are not used, as they become too diffusive and are not accurate enough.
At first order, the \textit{implicit} Runge--Kutta method reads
\begin{equation*}
  \mathrm{RK}_{\mathbf F}^{(1)} (\Delta t; \mathbf X ) = \mathbf F( t^{n}, \mathbf X ).
\end{equation*}
To evaluate $\mathbf F( t^n, \mathbf X )$ at any points $(z, \theta)$ in the phase-space, we need to interpolate them from the known grid values $\mathbf F_{ij}( t^n )$.
Historically, this scheme was first employed in \cite{PhysRevLett.119.220604} to solve equation~\eqref{eq:GHD_equation} in the absence of any external potential.
Similarly, we can construct an \textit{explicit} first-order Runge--Kutta scheme such as
\begin{equation*}
  \mathrm{RK}_{\mathbf F}^{(1)} (\Delta t;  \mathbf x_{ij} ) = \mathbf F( t^{n+1}, \mathbf x_{ij}),
\end{equation*}
where $\mathbf F( t^{n+1}, \mathbf x_{ij}) \equiv \mathbf F_{ij}( t^{n+1}) $ is estimated via equation~\eqref{eq:velocity_expansion}.

\subsubsection{Second-order schemes (RK2)}
At second order, the Runge--Kutta schemes employ the midpoint advection field $\mathbf F(t^{n+1/2})$ to obtain a more accurate estimate of the origin of the characteristics.
To estimate the midpoint advection field on the grid we employ the time-Taylor series expansion \eqref{eq:velocity_expansion}
and the recursive procedure of Section~\ref{s:ATDAF} to compute the  time derivatives  of the advection field $\{\partial_t^{(k)} \mathbf F(t)\}_{k\geq 0}$.
Next, we use cubic-spline interpolation to evaluate $\mathbf F( t^{n+1/2}, \mathbf X )$ at any points $(z, \theta)$ in the phase-space.
Integrating the ODEs between $t^n$ and $t^{n+1}$ with a second-order quadrature in time, we obtain the following  \textit{implicit} scheme,
\begin{equation*}
    \mathrm{RK}_{\mathbf F}^{(2)} (\Delta t; \mathbf X) = \mathbf F \bigg( t^{n + 1/2},  \frac{ \mathbf X + \mathbf X (t^{n+1}) }{2} \bigg),
\end{equation*}
where $( \mathbf X (t^{n}) + \mathbf X (t^{n+1}) ) /2 \simeq \mathbf X (t^{n+1/2}) + \mathcal{O}(\Delta t^2) $
is the second-order approximation of the midpoint characteristics.
The corresponding \textit{explicit} scheme reads
\begin{equation*}
    \mathrm{RK}_{\mathbf F}^{(2)} (\Delta t; \mathbf x_{ij}) = \mathbf F \Big( t^{n+1/2},  \mathbf x_{ij} - \frac{\Delta t}{2} \mathbf F_{ij}(t^{n})\Big).
\end{equation*}
Similar time discretizations were already used in \cite{XIU2001658}.

\subsubsection{Fourth-order schemes (RK4)}
\label{ss:RK4}
Assume that for all times $t \in [t^n,\, t^{n+1}]$, the approximate advection field $\mathbf F(t)$ is known. This can be obtained by using
the recursive method of Section~\ref{s:ATDAF}. We recall that this method allows us to compute recursively
the time derivatives  of the advection field $\{\partial_t^{(k)} \mathbf F(t)\}_{k\geq 0}$, in order to construct the time-Taylor series expansion \eqref{eq:velocity_expansion} of the advection field  $\mathbf F(t)$.
We then integrate the "backward" characteristic which ends at time $t^{n+1}$ at the grid point $x_{ij}$.
At fourth order the Runge--Kutta scheme then reads
\begin{equation*}
  \mathrm{RK}_{\mathbf F}^{(4)} (\Delta t; \mathbf X) = \frac{1}{6} (K_{1}+2 K_{2}+2 K_{3}+K_{4}),
\end{equation*}
where the \textit{implicit} advection estimates are given by
\begin{subequations}
  \begin{align}
    &K_{1}=\mathbf F(t^{n} , \mathbf X) \\
    &K_{2}=\mathbf F\Big(t^{n+1/2}, \mathbf X+\frac{\Delta t}{2} K_{1}\Big) \\
    &K_{3}=\mathbf F\Big(t^{n+1/2}, \mathbf X+\frac{\Delta t}{2} K_{2}\Big) \\
    &K_{4}=\mathbf F(t^{n+1}, \mathbf X +\Delta t K_{3}). \label{L4IRK4}
  \end{align}
\end{subequations}
Since the coordinate argument of $K_4$ in \eqref{L4IRK4}, namely $\mathbf X^{n}+\Delta t K_{3}$,
is supposed to be an estimate of the arrival point at $t^{n+1}$, which coincides with the grid as principle of the BSL scheme, we then replace
\eqref{L4IRK4} by $K_{4}=\mathbf F(t^{n+1}, \mathbf x_{ij})$.
Then, only $K_1$, $K_2$ and $K_3$  depend implicitly on the unknown starting point $\mathbf X$ at time $t^n$, so an implicit equation such as \eqref{eq:RK_fixedpoint} has to be solved to find this point by using the Picard or Newton iteration.
The implicit scheme can   be seen as integration of the "backward" characteristic in forward direction.
\\
In its \textit{explicit} form, the advection estimates take the following form,
\begin{equation}
\begin{aligned}
    &K_{1}=\mathbf F (t^{n} , \mathbf x_{ij} ) \\
    &K_{2}=\mathbf F\Big(t^{n+1/2}, \mathbf x_{ij}-\frac{\Delta t}{2} K_{1}\Big) \\
    &K_{3}=\mathbf F\Big(t^{n+1/2}, \mathbf x_{ij}-\frac{\Delta t}{2} K_{2}\Big) \\
    &K_{4}=\mathbf F(t^{n+1}, \mathbf x_{ij}-\Delta t K_{3}).
\end{aligned}
\label{ERK4}
\end{equation}
The explicit scheme can be seen as integrating the "backward" characteristic in backward direction, hence the negative time increment.
This method is a generalization to fourth order of the second-order scheme used in \cite{XIU2001658}.

\begin{remark}
  It should be noted that one can also construct a \textit{fully implicit} 
  2-stage
  Runge--Kutta scheme of fourth order
  following the method by Hammer and Hollingsworth~\cite{10.2307/2002064} where
  \begin{equation*}
    \mathrm{RK}_{\mathbf F}^{(4)} (\Delta t; \mathbf X) = b_1 \mathbf F (t^{n+1} - c_1 \Delta t, \mathbf x_{ij} - \Delta t K_1)
    + b_2 \mathbf F (t^{n+1} - c_2 \Delta t, \mathbf x_{ij} - \Delta t K_2),
  \end{equation*}
  and the fully implicit advection estimates take the following form,
  \begin{equation*}
    \begin{aligned}
      &K_{1}= a_{11} \mathbf F(t^{n+1} - c_1 \Delta t, \mathbf x_{ij} - \Delta t K_1 )
      + a_{12} \mathbf F(t^{n+1} - c_2 \Delta t, \mathbf x_{ij} - \Delta t K_2 ) \\
      &K_{2}= a_{21} \mathbf F(t^{n+1} - c_1 \Delta t, \mathbf x_{ij} - \Delta t K_1 )
      + a_{22} \mathbf F(t^{n+1} - c_2 \Delta t, \mathbf x_{ij} - \Delta t K_2 ).
    \end{aligned}
  \end{equation*}
  Here, the  coefficients $a_{ij}$ and $b_i$ are given by
  \begin{equation*}
    \begin{aligned}
      &a_{11} = \frac{1}{4}, \quad a_{12} = \frac{1}{4} - \frac{\sqrt{3}}{6}, \quad a_{21} = \frac{1}{4} + \frac{\sqrt{3}}{6} \;, \quad a_{22} = \frac{1}{4}, \\
      &b_1 = b_2 = \frac{1}{2}, \\
      &c_1 = \frac{1}{2} - \frac{\sqrt{3}}{6}, \quad c_2 = \frac{1}{2} + \frac{\sqrt{3}}{6}.
    \end{aligned}
  \end{equation*}
  We find that this fully implicit scheme offers a small performance increase over the previous implicit scheme.
\end{remark}

\subsection{Adams--Moulton semi-Lagrangian schemes}
\label{ss:AMSLS}
Adams--Moulton schemes are a family of \textit{implicit} schemes, which employ the solutions at previous time steps
to obtain a higher-order approximation for the solution of ordinary differential equations.
When applied to equation~\eqref{eq:BSL_ODE},  Adams--Moulton schemes require knowledge of the endpoint advection field $\mathbf F^{n+1}$. 
To this end, a high-order reconstruction of $\mathbf F^{n+1}$ can be obtained by using either an affine extrapolation in time using the
known advection field at previous times or by using the recursice procedure of Section~\ref{s:ATDAF}, which allows us to compute recursively
the time derivatives of the advection field, in order to construct the time-Taylor series expansion of the advection field \eqref{eq:velocity_expansion}. 
As we will demonstrate, using the recursive procedure of Section~\ref{s:ATDAF} generally produces more accurate results than time extrapolations. 
Using the Adams--Moulton schemes of order $s$, that we formally write as
$\mathrm{AM}_{\mathbf F}^{(s)} (\Delta t; t^{n}, t^{n-1}, \ldots ,  ; \mathbf X (t^{n}))$, we obtain
\begin{equation}
\begin{aligned}
    \mathbf X (t^{n+1})- \mathbf X (t^{n}) &=\int_{t^{n}}^{t^{n+1}} \mathbf F(s , \mathbf X(s)) \mathrm{d}s \\
    & \simeq \Delta t \: \mathrm{AM}_{\mathbf F}^{(s)}\big (\Delta t; t^{n+1},  t^{n}, t^{n-1}, \ldots ,  ; \mathbf X (t^{n})\big).
\end{aligned}
\label{eq:AM_scheme}
\end{equation}
As before, we have to solve the following fixed-point problem to obtain the departure points $\mathbf X (t^n)$,
\begin{equation}
    \mathbf X \to G(\mathbf X) \coloneqq \mathbf x_{ij} - \mathbf X - \Delta t \: \mathrm{AM}_{\mathbf F}^{(s)} (\Delta t; t^{n+1}, t^{n}, t^{n-1} , \ldots ; \mathbf X ).
    \label{eq:AM_fixedpoint}
\end{equation}
Again, the problem $G(\mathbf X) = 0$ is solved by using the Picard iteration.

\subsubsection{Second-order schemes (AM2)}
Assume that for all times $t^{m}$, $m \leq n$, the approximate solutions of the filling function $f^m$ and the corresponding advection field $\mathbf F^{m}$ are known. 
The second-order Adams--Moulton scheme reads~\cite{FILBET2016171},
\begin{equation*}
    \mathrm{AM}_{\mathbf F}^{(2)} (\Delta t; \mathbf X ) = \frac{1}{2} \big( \mathbf F( t^{n+1}, \mathbf x_{ij} ) + \mathbf F ( t^{n}, \mathbf X ) \big),
\end{equation*}
where $\mathbf F ( t^{n+1}, \mathbf x_{ij} )$ is obtained either via equation~\eqref{eq:velocity_expansion}, by
using the known grid values $ \{\partial_t^{(k)} \mathbf F_{ij}( t^n )\}_{k\geq 0}$, or through the second-order extrapolation in time,
\begin{equation*}
  \mathbf F ( t^{n+1}, \mathbf x_{ij} ) \simeq 2 \mathbf F ( t^{n}, \mathbf x_{ij} ) - \mathbf F( t^{n-1}, \mathbf x_{ij} ).
\end{equation*}

\subsubsection{Fourth-order schemes (AM4)}
Assume that for all times $t^{m}$, $m \leq n$, the approximate solutions of the filling function $f^m$
and the corresponding advection field $\mathbf F^{m}$ are known. 
The fourth-order Adams--Moulton scheme then reads~\cite{FILBET2016171},
\begin{equation*}
\begin{dcases}
  \mathrm{AM}_{\mathbf F}^{(4)} (\mathbf X ) & = \frac{1}{24} \big( 9 \mathbf F ( t^{n+1}, \mathbf x_{ij} ) + 19 \mathbf F ( t^{n}, \mathbf X )
  - 5 \mathbf F ( t^{n-1}, \mathbf X^{n-1}) + \mathbf F ( t^{n-2}, \mathbf X^{n-2} ) \big) \\
    \mathbf X^{n-1} & = \mathbf x_{ij} + 4 ( \mathbf x_{ij} -  \mathbf X ) -  2 \Delta t \big( 2 \mathbf F( t^n , \mathbf X) +  \mathbf F( t^{n+1}, \mathbf x_{ij}) \big) \\
    \mathbf X^{n-2} & = \mathbf x_{ij} + 27 ( \mathbf x_{ij} -  \mathbf X ) -  6 \Delta t \big( 3 \mathbf F ( t^n , \mathbf X) +
    2 \mathbf F ( t^{n+1}, \mathbf x_{ij} ) \big). 
\end{dcases}
\end{equation*}
The advection field $\mathbf F ( t^{n+1}, \mathbf x_{ij})$ is obtained either via equation~\eqref{eq:velocity_expansion}, by using the known grid values
$ \{\partial_t^{k} \mathbf F_{ij}(t^n)\}_{k \geq 0}$, or through the fourth-order extrapolation in time,
\begin{equation*}
  \mathbf F ( t^{n+1}, \mathbf x_{ij} ) \simeq 4 \mathbf F ( t^{n}, \mathbf x_{ij}) -
  6 \mathbf F ( t^{n-1}, \mathbf x_{ij} ) + 4 \mathbf F ( t^{n-2}, \mathbf x_{ij}) -  \mathbf F( t^{n-3}, \mathbf x_{ij}).
\end{equation*}

\subsection{Leap-frog schemes (LF2)}
\label{ss:LF}
Unlike the Runge--Kutta schemes, leap-frog methods do not estimate the midpoint advection field using a time-Taylor series  expansion.
Instead, they rely on taking intermediate steps of the function $f$, from which the advection field can be computed. 
Here, we present an implicit second-order leap-frog scheme based on the semi-Lagrangian technique~\cite{FILBET2016171}.
Assume that for all times $t^{m}$, $m \leq n$, the approximate solutions of the filling function $f^m$ and the corresponding advection field $\mathbf F^{m}$ are known. 
First, we estimate the midpoint solution $f^{n+1/2}$ by solving the fixed-point problem,
\begin{equation*}
\begin{dcases}
    \mathbf X^{n+1/2} - \mathbf X^n - \frac{\Delta t}{2} \mathbf F (t^n, \mathbf X^n)  = 0 \\
    \mathbf X^{n+1/2} = \mathbf x_{ij},
\end{dcases}
\end{equation*}
with the help of the Picard iteration and then interpolating the filling function at $\mathbf X^n$ to the mesh points $\mathbf x_{ij}$, namely,
\begin{equation*}
    f^{n+1/2}(\mathbf{x}_{i})= \Pi_{h} f^{n}(\mathbf{X}^{n}).
\end{equation*}
Next, from $f^{n+1/2}$, the advection field $\mathbf F^{n+1/2}$ is calculated, and then used to compute a new approximation
of $\mathbf X^n$ by solving the fixed-point problem,
\begin{equation*}
\begin{dcases}
    \mathbf X^{n+1} - \mathbf X^n - \Delta t \mathbf F \Big( t^{n+1/2}, \frac{ \mathbf X^{n+1} + \mathbf X^{n} } {2} \Big) = 0 \\
    \mathbf X^{n+1} = \mathbf x_{ij}.
\end{dcases}
\end{equation*}
Finally, the solution at time $t^{n+1}$ is given through interpolation, i.e. 
\begin{equation*}
  f^{n+1}(\mathbf{x}_{i})= \Pi_{h} f^{n}(\mathbf{X}^{n}).
\end{equation*}
An explicit form of the second-order leap-frog scheme for solving the GHD advection equation was proposed in \cite{PhysRevLett.123.130602}.
This scheme requires knowledge of the approximate solutions of the filling function both at time $t^{n}$ and $t^{n+1/2}$.
Given the latter, the advection field $\mathbf F^{n+1/2}$ is calculated and the departure points $\mathbf X^n$ are obtained from the following scheme,
\begin{equation}
\begin{dcases}
    \mathbf X^n & = \mathbf x_{ij} - \Delta t \mathbf F ( t^{n+1/2}, \mathbf X^{n+1/2} ) \\
    \mathbf X^{n+1/2} & = \mathbf x_{ij} - \frac{\Delta t}{2} \mathbf F ( t^{n+1/2}, \mathbf x_{ij} ).
\end{dcases}
\label{eq:LF2_explicit}
\end{equation}
Hence, the solution at time $t^{n+1}$ is given through interpolation by
\begin{equation}
    f^{n+1}(\mathbf{x}_{i})= \Pi_{h} f^{n}(\mathbf{X}^{n}).
    \label{eq:LF2_int}
\end{equation}
To obtain the solution at time $t^{n+3/2}$, required for the next iteration, one must repeat the step described by
equations~\eqref{eq:LF2_explicit} and \eqref{eq:LF2_int} with the newly obtained solution $f^{n+1}$ replacing $f^{n+1/2}$.

\subsection{BSL-hybrid schemes for dissipative GHD}
\label{ss:BSLH}
Given the departure points $\mathbf X^n$ found using BSL schemes for the GHD advection equation~\eqref{eq:GHD_equation},
solving the extended GHD equation~\eqref{eq:GHD_equation_nonlin} for a single time step requires integrating
the nonlinear source term along the characteristics, namely
\begin{equation}
  f^{n+1} (\mathbf x_{ij}) = f^{n} (\mathbf X^n) + \int_{t^n}^{t^{n+1}} \mathrm{d}\tau \: \mathcal{N}(\tau,\mathbf{X}(\tau),f(\tau)).
  \label{eq:nonlin_source_integral}
\end{equation}
To this end, various time discretization methods can be employed to create BSL-hybrid schemes.
In the following we will review a few such schemes.

\subsubsection{The endpoint scheme}
The simplest scheme consists of estimating the integral~\eqref{eq:nonlin_source_integral} by its endpoint, giving
\begin{equation*}
  f^{n+1} (\mathbf x_{ij}) - f^n (\mathbf X^n) = \Delta t \mathcal{N}(t^n, \mathbf X^n).
\end{equation*}

\subsubsection{The midpoint scheme}
Alternatively, one can employ the source term at the midpoint of the time step to obtain,
\begin{equation}
  f^{n+1} (\mathbf x_{ij}) - f^n (\mathbf X^n) = \frac{\Delta t}{2} \big( \mathcal{N} (t^{n+1/2} ,
  \mathbf x_{ij}) + \mathcal{N} (t^{n+1/2} ,\mathbf X^n)  \big).
  \label{def:midpoint}
\end{equation}
Here, the midpoint along the trajectory is obtained by averaging the source term at its start and endpoint.
This procedure improves the stability, yet retains the second-order accuracy, of the more standard scheme, which consists
in evaluating $\mathcal{N} (t^{n+1/2})$ at the middle of the trajectory $(\mathbf x_{ij}+\mathbf X^n)/2$, i.e.
by substituting $\Delta t\mathcal{N} (t^{n+1/2},(\mathbf x_{ij}+\mathbf X^n)/2)$ to right-hand side of \eqref{def:midpoint} (see \cite{Hor02}). 
The midpoint in time is approximated by linear extrapolation from the previous time step such as
\begin{equation*}
    \mathcal{N} (t^{n+1/2} ,\mathbf X) \simeq \frac{3}{2} \mathcal{N} (t^n , \mathbf X) - \frac{1}{2} \mathcal{N} (t^{n-1}, \mathbf X).
\end{equation*}
Higher-order estimates of the temporal midpoint can be obtained by including additional previous time steps,
however, such extrapolations may also be more prone to noise \cite{Hor02}.

\subsubsection{The two-time-level semi-Lagrangian scheme (SETTLS)}
The extrapolation of the source term employed in the midpoint scheme can suffer from noise problems.
An alternative scheme is the SETTLS scheme~\cite{alma9918447213902959, Hor02, PS19}, given by
\begin{equation*}
  f^{n+1} (\mathbf x_{ij}) - f^n (\mathbf X^n) = \frac{\Delta t}{2} \big( \mathcal{N} (t^{n} , \mathbf x_{ij}) + \mathcal{N} (t^{n+1} ,\mathbf X^n)  \big).
\end{equation*}
Here,  $\mathcal{N} (t^{n+1})$ is again approximated by linear extrapolation from the previous time step such as
\begin{equation*}
  \mathcal{N} (t^{n+1} ,\mathbf X) \simeq 2 \mathcal{N} (t^n , \mathbf X) -  \mathcal{N} (t^{n-1}, \mathbf X).
\end{equation*}

\subsubsection{The Crank--Nicolson scheme}
Applying a trapezoidal quadrature rule to estimate the integral~\eqref{eq:nonlin_source_integral} yields \cite{SK06},
\begin{equation}
  f^{n+1} (\mathbf x_{ij}) - f^n (\mathbf X^n) = \frac{\Delta t}{2} \big( \mathcal{N} (t^{n+1} , \mathbf x_{ij})
  + \mathcal{N} (t^{n} ,\mathbf X^n)  \big).
  \label{eq:nonlin_crank_nicolson}
\end{equation}
Unlike the previous methods, the Crank--Nicolson method 
is implicit as the source term $\mathcal{N} (t^{n+1} , \mathbf x_{ij})$ depends on the unknown $f^{n+1} (\mathbf x_{ij})$.
The problem~\eqref{eq:nonlin_crank_nicolson} can be solved by using the Picard iteration.
Note that repeatedly evaluating the source term is extremely costly. Therefore,
in the benchmarks below, the tolerance for the convergence of the Picard iteration is set so that the scheme converges in few iterations,
typically two or three.

\begin{remark}
For the inclusion of diffusion terms in \eqref{eq:GHD_equation_nonlin},
care must be taken of the fact that numerical diffusion results from the discretization of the advective part
of the equation. This error is usually of the order of the applied method, so that a discretization of the
diffusive term at the same order of accuracy is perturbed by a diffusion error at the same order. For this reason it is necessary to resolve the advection step with a method of higher order than the diffusive step.
In our case that means it suffices to consider second order methods for the diffusion step.
\end{remark}

\section{Benchmark of BSL and BSL-hybrid schemes for solving the dissipative GHD equations} \label{s:benchmarks}

To benchmark the performance of the various schemes, we will examine the evolution of three particular conserved quantities, namely the particle number $N$, the energy $E$, and the entropy $S$ given by the expressions
\begin{align}
  N (t) &=  \frac{1}{2 \pi} \int_\R \mathrm{d}z \int_\R \mathrm{d}\theta \:  1^{\mathrm{dr}}(t, z,\theta) f(t, z,\theta), \label{eq:atomnumber} \\
  E (t) &=   \frac{1}{2 \pi}\int_\R \mathrm{d}z \int_\R \mathrm{d}\theta \:
  \big( \theta^2 + V(z) \big) 1^{\mathrm{dr}}(t, z,\theta) f(t, z,\theta), \label{eq:energy} \\
  S(t) &= \frac{1}{2 \pi}\int_\R \mathrm{d}z \int_\R \mathrm{d}\theta \: 1^{\mathrm{dr}}(t, z,\theta) f^2(t, z,\theta). \label{eq:L2norm}
\end{align}
Note, the latter quantity corresponds to the entropy defined by \eqref{CL:S} with $\Psi(f)=f$ and is particularly sensitive to (numerical) diffusion.
To quantify the accuracy of each scheme, we compute the relative error of the conserved quantities w.r.t. their initial value, namely
\begin{equation*}
    \mathcal{E}_Q (t) = \frac{| Q(t) - Q(0) |}{Q(0)},
\end{equation*}
where $Q(t) = \{ N(t), E(t), S(t) \}$.

\begin{figure}[h]
    \centering
    \includegraphics{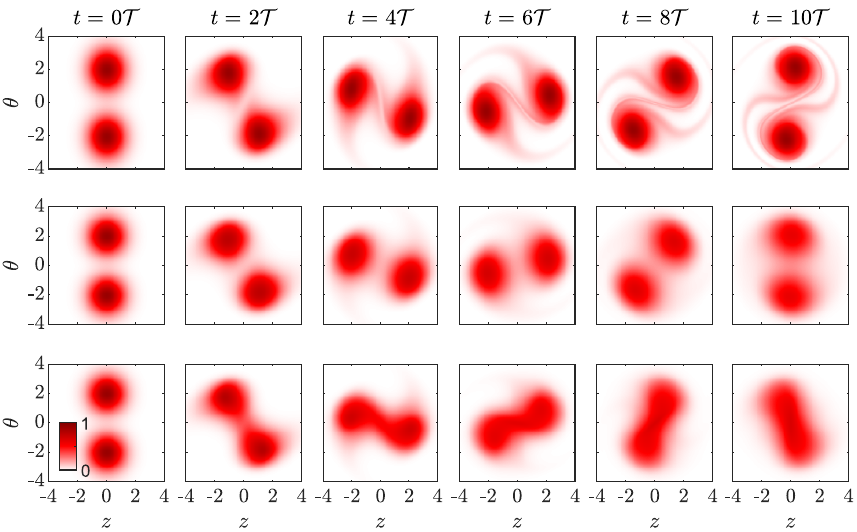}
    \caption{\label{fig:reference_solution}
    Example of filling function $f$ calculated for the quantum Newton's cradle benchmark problem.
    The initial state~\eqref{eq:benchmark_initial_state} is propagated for 10 oscillation periods $\mathcal{T}$ of the cradle following BSL (hybrid) methods.
    \textbf{Top row:} Purely advective GHD calculation performed using explicit Runge--Kutta scheme of fourth order (RK4) with equally-spaced grids $N_\theta=N_z=513$ and $N_{\mathrm{steps}} = 2000$ time steps.
    Note all solutions for $f$ obtained using higher-order methods appear visually very similar. 
    \textbf{Middle and bottom rows:} Dissipative GHD calculations performed using BSL hybrid schemes (SETTLS + explicit RK4) with equally-spaced grids $N_\theta=N_z=257$ and $N_{\mathrm{steps}} = 3000$ time steps.
    The dissipative GHD equation is solved for the diffusion operator (middle row) and the Boltzmann-like collision integral (bottom row).  
    Note the grids employed for the computation extend further than shown in the figure.
    Further, all solutions obtained for $f$ appear visually very similar.
    }
\end{figure}

For the benchmark, we will solve the ODE~\eqref{eq:BSL_ODE} for a relatively difficult problem, namely the Quantum Newton's
cradle setup~\cite{kinoshita2006quantum, 10.21468/SciPostPhys.9.4.058, PhysRevLett.126.090602}.
The setup can be realized experimentally using ultracold atomic gases and constitutes one of the most remarkable demonstrations of integrability. 
In the Quantum Newton's cradle setup, the system is initialized in a (nearly) harmonic, external potential $V(z)$.
Next, two large, opposite momenta are imparted onto the system, causing the initial (typically thermal) state to split into two atomic clouds oscillating in the external confinement.
Every half-period of the cradle oscillation the clouds collide in the center of the potential.
The ensuing interactions, encoded in the effective velocity, deform the clouds.
Further, a small anharmonicity of the external potential can lead to the formation of very fine structures in the filling function,
thus demanding high accuracy of the numerical solutions~\cite{10.21468/SciPostPhys.6.6.070}.
An example of this can be seen in Fig.~\ref{fig:reference_solution}.

In GHD simulations of experimental systems, the initial filling function $f_0$ is often given by a thermal state obtained via the thermodynamic Bethe Ansatz~\cite{YY1}.
However, for easier reproducibility, we instead let it be given by the expression
\begin{equation}
  f_0 (z,\theta) = 0.9 \left[ \exp \left( - \frac{\left( \theta - \theta_\mathrm{B} \right)^2 }{2 \sigma} \right)
    + \exp \left( - \frac{\left( \theta + \theta_\mathrm{B} \right)^2 }{2 \sigma} \right) \right] \exp \left( -\frac{ z ^2 }{2 \sigma} \right),
    \label{eq:benchmark_initial_state}
\end{equation}
where $\theta_\mathrm{{B}} = 2$ and $\sigma = 1/\sqrt{2}$.
Further, we employ an external potential in the shape of a Gaussian, imitating the intensity profile of a laser beam
typically used to generate the potential in an experimental setting
\begin{equation*}
    V(z) = \frac{1}{8}\omega^2 \eta^2 \left[ 1 - \exp \left( - \frac{2 z^2}{ \eta^2} \right) \right].
\end{equation*}
For the benchmarks we set $\omega = 2$ and $\eta = 12$.
The coupling constant is set to $c=1$.
We will be expressing time in units of the oscillation period
$\mathcal{T} = 2 \pi / \omega$, and the simulations will be carried out for a duration of 10 periods.
Whenever the time-Taylor expansion~\eqref{eq:velocity_expansion} is used to estimate the advection fields at future times, only the first order time derivatives of the advection fields are employed.
All numerical schemes for solving the GHD equations are implemented using the iFluid framework~\cite{10.21468/SciPostPhys.8.3.041}.

\subsection{BSL schemes for solving the advective GHD equation}

For the benchmarks of the BSL schemes for solving the advective GHD equation, we employ uniformly spaced position and rapidity grids discretized into $N_\theta=N_z=513$ grid points.
Such high resolution is necessary to properly resolve the fine structures appearing in the filling function $f$ following dynamics (see Fig.~\ref{fig:reference_solution}).
We test the accuracy of the various schemes for different time step resolutions, discretizing the evolution duration into $N_{\mathrm{steps}} = 500$, 1000, and 2000 time steps.

However, before comparing their accuracy, it is worth discussing the numerical cost of performing a single time step following the different schemes. 
Among all operations needed for performing a BSL time step of the GHD equation, the most numerically expensive one is, by far, solving the dressing equation~\eqref{eq:dressing_linear}.
Calculating the effective velocity requires solving the dressing equation for both the rapidity derivative of the single-particle energy $\partial_\theta \epsilon (\theta)$ and momentum $\partial_\theta p (\theta)$.
Further, if the time-Taylor series expansion~\eqref{eq:velocity_expansion} is used to compute the advection fields at future times, additional dressing equations must be solved, as calculating the (first order) time derivatives of the advection fields requires dressing the source terms $S_w$ and $S_b$.
Note, for the Lieb-Liniger model whose single-particle momentum reads $p(\theta) = \theta$, solving the dressing equation for the unit-cell $1^{\mathrm{dr}}$ is not necessary, as it is already given by the dressed rapidity derivative of the single-particle momentum.

When timing the different schemes, we find that the total cost of performing a single BSL step is indeed completely dominated by solving the dressing equation.
In the implicit schemes, the advection fields (and their time derivatives) are calculated only once at each propagation step and then interpolated to the solution $\mathbf X^n$ at each iteration.
In comparison to computing the fields, performing the phase space interpolation is a relatively cheap operation, resulting in a rather small difference in runtime between implicit and explicit schemes.
Restricting ourselves to first-order time derivatives of the advection fields, we find the propagation step of higher-order Runge--Kutta schemes (RK2 and RK4) taking roughly twice the time to execute than the first-order scheme (RK1), as the latter does not employ estimates of the advection fields at intermediate times.
Meanwhile, the leap-frog schemes perform an additional, intermediate step for every full time step, thus having to evaluate the effective velocity twice.
Hence, for a full $\Delta t$ time step, the leap-frog schemes exhibit computation times similar to the higher-order Runge--Kutta schemes.
Lastly, the Adams--Moulton schemes employing the time-Taylor series expansion show similar computational cost to the higher-order Runge--Kutta schemes.
However, when employing an extrapolation of the advection fields using previous steps, the Adams--Moulton schemes obtain a numerical efficiency compared to the RK1 schemes, since only a single evaluation of the advection fields is needed.

Note that all these observations are specific to the problem of solving the GHD equation. In most other problems, computing the advection fields does not represent such a significant fraction of the computational load, whereby a discernible difference in the runtime between implicit and explicit schemes would be found.
Furthermore, for problems involving higher-dimensional systems, storing the full solution at multiple time-steps (needed for the Adams--Moulton schemes) may introduce significant overhead.

\begin{figure}[h]
    \centering
    \includegraphics{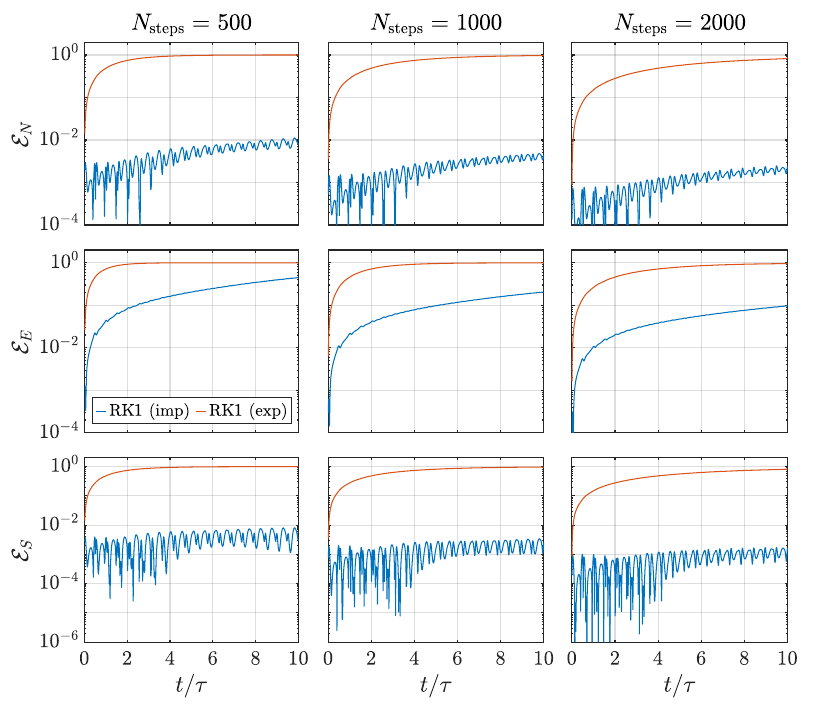}
    \caption{\label{fig:BSL_benchmark_order1}
    Benchmark of first-order BSL schemes for the quantum Newton's cradle problem.
    In the legend, the labels "imp" and "exp" indicate that the methods are implicit or explicit, respectively.
    The calculation is performed on equally-spaced grids with resolutions $N_\theta=N_z=513$ and for various number of time steps $N_{\mathrm{steps}}$.
    Plotted is the relative error of the particles number, total energy, and entropy.
    }
\end{figure}

\begin{figure}[h]
    \centering
    \includegraphics{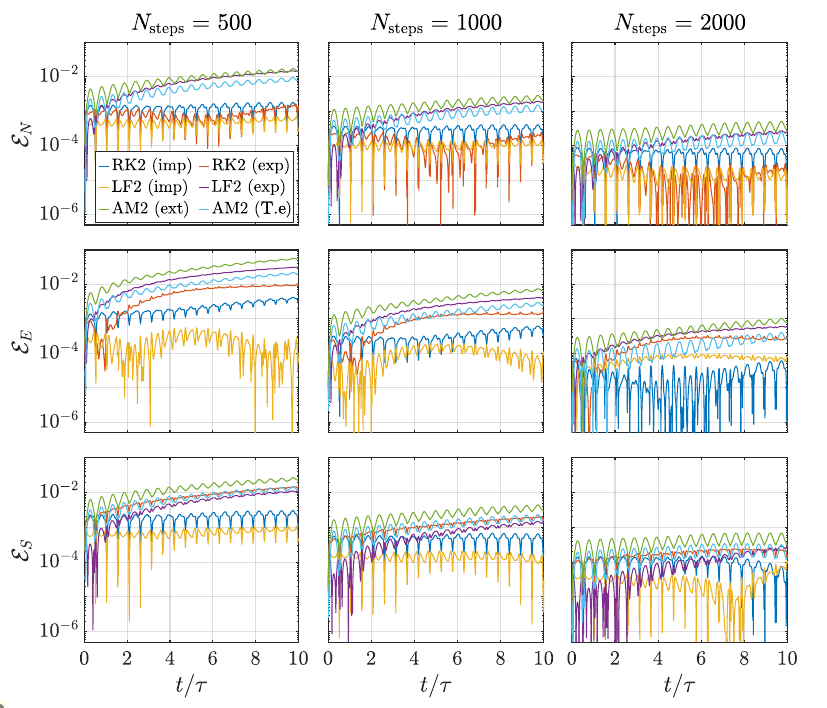}
    \caption{\label{fig:BSL_benchmark_order2}
    Benchmark of second-order BSL schemes for the quantum Newton's cradle problem.
    In the legend, the labels "imp" and "exp" indicate that the methods are implicit or explicit, respectively.
    For the Adams--Moulton schemes (AM), the label "ext" denotes that the advection fields at the endpoint are obtained using extrapolation from previous time steps.
    Meanwhile, "T.e" indicates that the time-Taylor expansion series has been employed.
    The calculation is performed on equally-spaced grids with resolutions $N_\theta=N_z=513$ and for various number of time steps $N_{\mathrm{steps}}$.
    Plotted is the relative error of the particles number, total energy, and entropy.
    }
\end{figure}

\begin{figure}[h]
    \centering
    \includegraphics{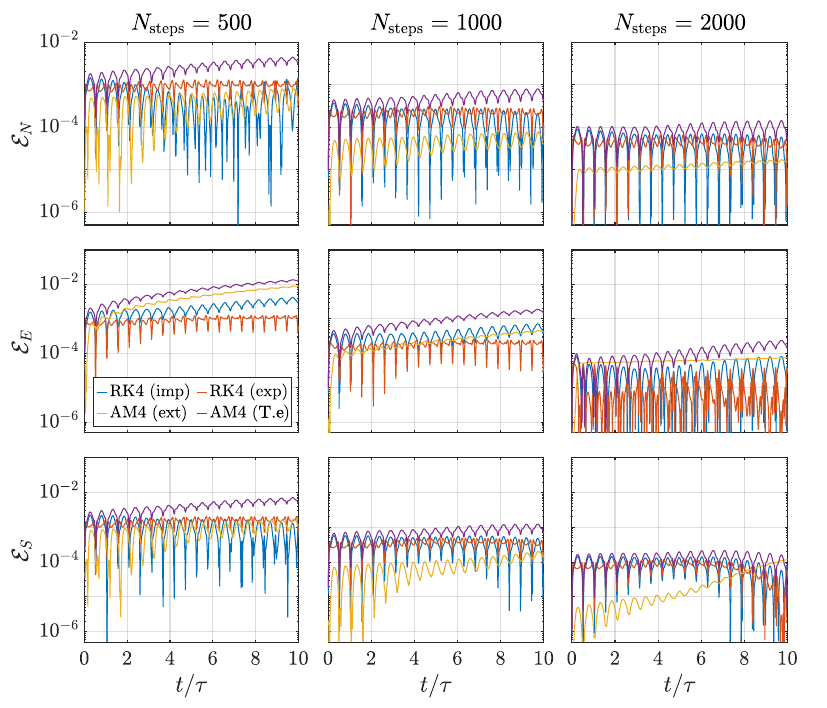}
    \caption{\label{fig:BSL_benchmark_order4}
    Benchmark of fourth-order BSL schemes for the quantum Newton's cradle problem.
    In the legend, the labels "imp" and "exp" indicate that the methods are implicit or explicit, respectively.
    For the Adams--Moulton schemes (AM), the label "ext" denotes that the advection fields at the endpoint are obtained using extrapolation from previous time steps.
    Meanwhile, "T.e" indicates that the time-Taylor expansion series has been employed.
    The calculation is performed on equally-spaced grids with resolutions $N_\theta=N_z=513$ and for various number of time steps $N_{\mathrm{steps}}$.
    Plotted is the relative error of the particles number, total energy, and entropy.
    }
\end{figure}

In Figures \ref{fig:BSL_benchmark_order1}, \ref{fig:BSL_benchmark_order2}, and \ref{fig:BSL_benchmark_order4} we plot the benchmark results for the first order, second order, and fourth order BSL schemes, respectively.

Starting with the first order schemes of Figure~\ref{fig:BSL_benchmark_order1}, we find that the explicit form of the RK1 algorithm performs extremely poorly, which is, perhaps, to be expected.
In comparison, the implicit form yields much more accurate results, although the relative change in the conserved quantities is still fairly large.
For the implicit scheme, we find the error of both the particle number and the entropy to exhibit rapid oscillations at a frequency corresponding to roughly half the period of the quantum Newton's cradle.
The oscillation frequency coincides with the overlap of the two atomic clouds in the center of the external trap, where interactions are much more significant than throughout the rest of the evolution.
Therefore, as the clouds overlap in position, the numerical solution becomes more sensitive to errors and additional precision is often needed.

Compared to the first order schemes, the second order schemes, whose benchmarks are plotted in Figure~\ref{fig:BSL_benchmark_order2}, are much more accurate.
Comparing the implicit versus explicit schemes for the Runge--Kutta and leap-frog methods, we generally find the implicit schemes to perform better.
In particular the implicit leap-frog scheme exhibits very low relative errors of all three conserved quantities calculated.
Furthermore, the magnitudes of its errors remain fairly stable throughout the whole evolution duration, as opposed to most of the other schemes, whose associated errors increase steadily.
Lastly, both of the two Adams--Moulton schemes exhibit worse performance than the Runge--Kutta schemes of same order.
Between the two of them, the scheme employing the time-Taylor series expansion~\eqref{eq:velocity_expansion}, as opposed to the linear extrapolation from previous time steps, to obtain the advection fields at time $t^{n+1}$ is more accurate.
However, when solving the GHD equation, the latter is much cheaper to perform numerically.
Interestingly, all the implicit schemes tested exhibit the same oscillating behaviour in the errors of the benchmark quantities, while the explicit schemes appear much more stable in this regard.

Finally, in Figure~\ref{fig:BSL_benchmark_order4} we compare the different fourth-order schemes.
In general, all schemes exhibit very high accuracy.
For increasing number of time steps, we observe increasing accuracy and thus convergence of the schemes. 
At fixed number of time steps, we observe that both of the 
Runge--Kutta
schemes perform better than the Adams--Moulton schemes.
Contrary to the second-order schemes, the explicit Runge--Kutta scheme now performs better than its implicit counterpart, and extrapolating of the advection fields from previous time steps yields better results for the the Adams--Moulton scheme than employing the time-Taylor series expansion~\eqref{eq:velocity_expansion}.
The improvement of the explicit scheme is most likely due to the trick that we discovered, where the initial advection estimate of the Runge--Kutta method $K_1$ is computed using the known state at time $t^n$.
Meanwhile, the worse performance of the Adams--Moulton scheme with time-Taylor series expanded advection fields might be due to the limited phase-space resolution.
Indeed, when computing the successive time derivatives of the advection field $\mathbf F$ according to the recursive procedure of Section~\ref{s:ATDAF}, derivatives in the position and rapidity variables of the filling function $f$ are required.
Here, the phase-space derivatives of $f$ are obtained through finite-difference methods.
At high enough time resolution, the limiting factor of the time-Taylor series expansion scheme may become the error of said finite-difference derivatives together with the limited phase-space resolution of the grid.

\subsection{BSL-hybrid schemes for solving the dissipative GHD equations}

For the benchmarks of the BSL-hybrid schemes for solving the dissipative GHD equations, we employ uniformly spaced position and rapidity grids discretized into $N_\theta=N_z=257$ grid points.
Compared to the purely advective case, coarser grids can typically be employed when accounting for dissipation, as these effects tend to smooth out the filling function $f$ following dynamics.
We test two dissipative nonlinear source terms, namely the diffusion operator~\eqref{eq:diffusion_kernel}
and the Boltzmann-type collision integral \eqref{eq:collision_integral} governing transverse state-changing collisions.
The two source terms have very different structures.
The diffusion operator scales with the position gradient of the filling function, thus leading to a smearing of its edges.
Meanwhile, the transverse excitations effectively destroy quasi-particles at high rapidities and re-create them at low rapidities, while de-excitations from transverse excited states lead to a general redistribution of quasi-particles across the ($z, \theta$) phase-space.
Hence, although both source terms lead to eventual thermalization of the system, the exact mechanisms, and thus dynamics towards thermalization, differ.
Examples of this can be seen in Fig.~\ref{fig:reference_solution}, where solutions for the filling function $f$ for the two different source-terms are plotted.

We test the accuracy of the various schemes for different time step resolutions, discretizing the evolution duration into $N_{\mathrm{steps}} = 2000$, 2500, and 3000 time steps.
To compute the departure points of the characteristics, the explicit form of the Runge--Kutta scheme of fourth order (RK4) is employed.
In addition to computing the error of the particle number, total energy, and entropy, we also calculate and plot the minimum value of the filling function $f$.

Compared to the pure BSL schemes, each time step of the BSL-hybrid schemes is much more numerically costly to perform due to the evaluation of the source terms.
Hence, the endpoint, midpoint and SETTLS schemes all exhibit very similar runtimes, as the computation is completely dominated by the single calculation of the source term needed.
Conversely, following the Crank-Nicolson scheme, the source term must be calculated anew at every iteration, making this scheme extremely expensive compared to the others.
In order to keep the computational runtime at a reasonable limit, we set the limit of convergence for the Crank-Nicolson scheme at a level attainable after just 2-3 iterations.


\subsubsection{Diffusion}

\begin{figure}[h]
    \centering
    \includegraphics{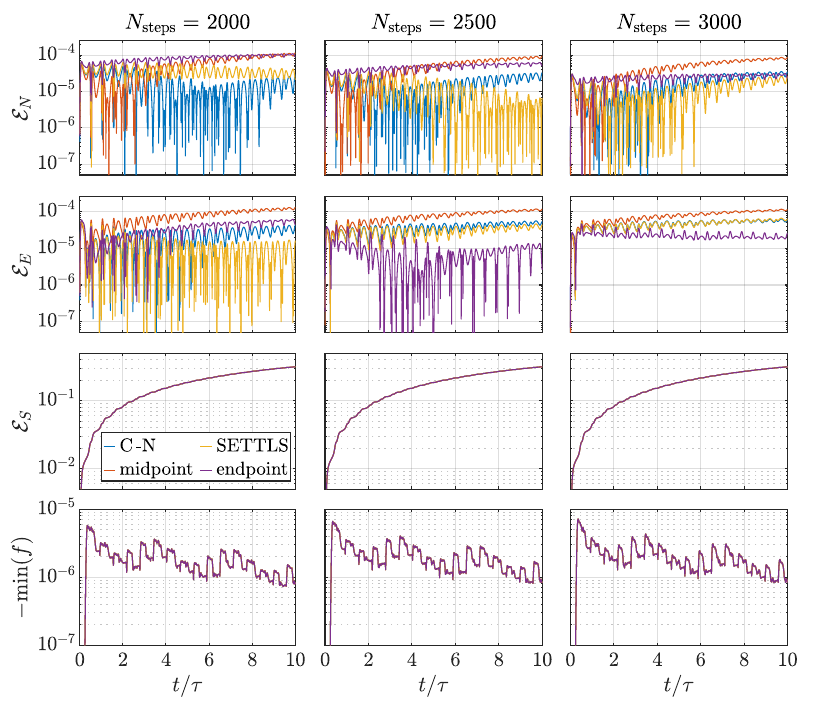}
    \caption{\label{fig:diffusion_benchmark}
    Benchmark of BSL-hybrid schemes for solving the dissipative GHD equation, here with the diffusion kernel as the nonlinear dissipative term.
    The benchmarks are performed on quantum Newton's cradle problem using the explicit RK4 method.
    The calculation is performed on equally-spaced grids with resolutions $N_\theta=N_z=257$ and for various number of time steps $N_{\mathrm{steps}}$.
    Plotted is the relative error of the particle number, total energy, and entropy, as well as the negative minimum value of the filling function $f$.
    }
\end{figure}

Figure~\ref{fig:diffusion_benchmark} shows the benchmark results of solving the quantum Newton's cradle problem accounting for diffusive effects.
Comparing to the previous benchmarks of the explicit RK4 method for solving the purely advective GHD equation, we do not find any significant increase in the relative error of neither particle number $N$ nor total energy $E$ when accounting for diffusion.
Meanwhile, the entropy increases significantly following diffusive dynamics, as expected.
Lastly, we do observe that the filling function $f$ at some points of the phase-space assumes negative values.
Recall that the filling function represents the fraction of allowed local momentum states occupied.
Hence, it should only assume values between 0 and 1.
However, the magnitude of the negative values found following numerical propagation is very small. 
Comparing the different hybrid schemes, not one single scheme consistently stands out in terms of accuracy, although for diffusive dynamics the midpoint scheme consistently appears slightly worse than the others.

\subsubsection{Boltzmann-type collision integral}

\begin{figure}[h]
    \centering
    \includegraphics{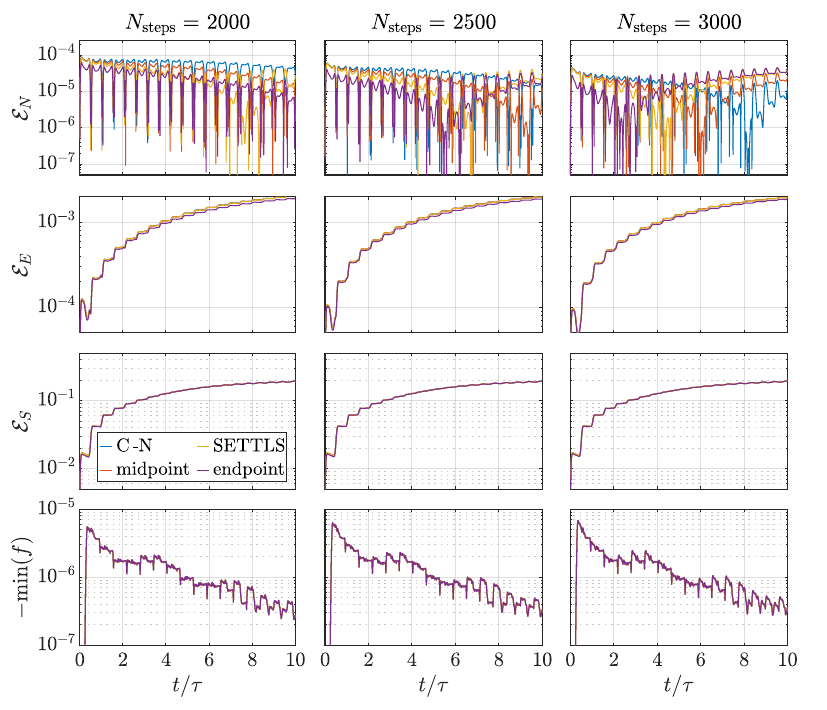}
    \caption{\label{fig:collision_benchmark}
    Benchmark of BSL-hybrid schemes for solving the dissipative GHD equation, here with the Boltzmann-type collision integral for transverse state-changing collisions as the nonlinear dissipative term.
    The benchmarks are performed on quantum Newton's cradle problem using the explicit RK4 method to obtain estimates of characteristics.
    The calculation is performed on equally-spaced grids with resolutions $N_\theta=N_z=257$ and for various number of time steps $N_{\mathrm{steps}}$.
    Plotted is the relative error of the particle number, total energy, and entropy, as well as the negative minimum value of the filling function $f$.
    }
\end{figure}

Unlike the case of diffusion (and most other instances of Boltzmann-type collision integrals), accounting for transverse excitations through a collision integral requires solving an additional set of differential equations, namely the rate equations for the excitation probabilities $\nu_n$~\eqref{eq:excitation_fraction}.
For the endpoint scheme, eq.~\eqref{eq:excitation_fraction} is numerically solved as follows
\begin{equation}
    \nu_n (t^{n+1}) = \nu_n (t^n) + \Delta t  \frac{1}{2} \beta_n \big[ \Gamma _{\mathrm{h}}^+ (t^{n}) - \Gamma _{\mathrm{p}}^+ (t^{n}) \nu_{n}^{\beta_n} (t^{n}) \big] \; ,
    \label{eq:nu_ODE_endpoint}
\end{equation}
while the remaining schemes employ midpoint estimates, such that
\begin{equation}
    \nu_n (t^{n+1}) = \nu_n (t^n) + \Delta t  \frac{1}{2} \beta_n \big[ \Gamma _{\mathrm{h}}^+ (t^{n+1/2}) - \Gamma _{\mathrm{p}}^+ (t^{n+1/2}) \: \nu_{n}^{\beta_n} (t^{n+1/2}) \big],
    \label{eq:nu_ODE_midpoint}
\end{equation}
where the function values at time $t^{n+1/2} = t^n + \Delta t /2\,$ are obtained by using the time extrapolation scheme
\begin{equation}
    \Gamma _{\mathrm{h}}^+ (t^{n+1/2}) = \frac{3}{2} \Gamma _{\mathrm{h}}^+ (t^{n}) - \frac{1}{2} \Gamma _{\mathrm{h}}^+ (t^{n-1}) \; .
\end{equation}
When accounting for atoms excited to transverse states beyond the ground state, we must also consider the associated transverse potential energy of the atoms can computing the total energy~\eqref{eq:energy}.
An atom in the $n$'th transverse state (where $n=0$ denotes the transverse ground state), has the transverse potential energy $2 n l_\perp^{-2}$ (in units $\hbar = 2m = 1$).
Thus, we can account for the transverse potential energy by adding the term $\sum_n \nu_n 2 n l_\perp^{-2}$ to the external potential $V(z)$ of eq.~\eqref{eq:energy}.
For these benchmarks we set the parameter $l_\perp = 1$.

Because the collision integral is originally constructed in the conservative form, we here employ a slightly different BSL-hybrid scheme:
First, the advective GHD equation is solved using a BSL scheme to obtain the characteristic origin.
Next, the filling function $f$ is transformed into the quasi-particle distribution $\rho_\mathrm{p}$, from which the collision integral is computed and accounted for.
Finally, the resulting quasi-particle distribution is transformed back to the filling function, marking the end of a single propagation step.\\

Figure~\ref{fig:collision_benchmark} shows the benchmark results of solving the quantum Newton's cradle problem accounting for transverse state-changing collisions via the collision integral.
Comparing to the previous benchmarks of the explicit RK4 method for solving the purely advective GHD equation, we do not find any significant increase in the relative error of neither particle number $N$ when accounting for transverse state-changing collisions.
Meanwhile, for all the schemes we find a very similar increase in the error of the total energy over time.
As already discussed, simplifications made in the construction of the collision integral result in the energy no longer being explicitly conserved.
The similarity of the error across all the tested schemes indicates that the source of the error is indeed due to the model, not the numerical solution.
Furthermore, we argued that the error associated with the simplifications would be rather small in most cases.
Indeed, the relative error of the total energy seen in Figure~\ref{fig:collision_benchmark} is only on the order of $0.1\%$ following 10 oscillation periods of the Newton's cradle.
Lastly, we again observe an increase in the error of entropy, as expected, and the filling function $f$ assuming negative values at some points of the $(z, \theta)$ phase-space.
However, again the magnitude of the negative values is very small for all the schemes.

\subsection{Summary of benchmarks}

\begin{table}[]
\centering
\renewcommand{\arraystretch}{1.2}
\begin{tabular}{l|llll}
\hline
\textbf{Numerical scheme}              &  $\int \varepsilon_{\mathrm{N}} \; \mathrm{d}\tau$ & $\int \varepsilon_{\mathrm{E}} \;\mathrm{d}\tau$        & $\int \varepsilon_{\mathrm{S}} \;\mathrm{d}\tau$        & Runtime  \\ \hline
\textbf{Advective GHD \eqref{eq:GHD_equation}} & \\ \hline
RK1 (implicit)      & 6.2e-3 & 2.9e-1 & 3.6e-3 & 1.1 \\
RK1 (explicit)      & 2.5e0 & 3.0e0 & 2.5e0 & \textbf{1.0}   \\
RK2 (implicit)      & 1.8e-4 & 1.2e-4 & 1.7e-4 & 2.1 \\
RK2 (explicit)      & 6.2e-5 & 8.3e-4 & 6.7e-4 & 2.0   \\
RK4 (implicit)      & 9.8e-5 & 1.4e-4 & 1.1e-4 & 2.2 \\
RK4 (explicit)      & 1.1e-4 & \textbf{2.5e-5} & \textbf{6.1e-5} & 2.1 \\
LF2 (implicit)      & \textbf{4.2e-5} & 2.2e-4 & 1.9e-4 & 2.1 \\
LF2 (explicit)      & 7.6e-4 & 1.8e-3 & 7.8e-4 & 2.0   \\
AM2 (extrapolation) & 1.2e-3 & 2.7e-3 & 1.6e-3 & 1.1 \\
AM2 (expansion)     & 5.5e-4 & 9.9e-4 & 6.8e-4 & 2.1 \\
AM4 (extrapolation) & 4.9e-5 & 2.3e-4 & 3.2e-4 & 1.2 \\
AM4 (expansion)     & 3.3e-4 & 6.1e-4 & 3.4e-4 & 2.2 \\ \hline
\textbf{Dissipative GHD \eqref{eq:GHD_equation_nonlin}: Collisions} & & & & \\ \hline
Crank-Nicolson & \textbf{2.7e-5} & 6.1e-3 & 5.9e-1 & 7.2 \\
midpoint       & 6.7e-5 & 6.0e-3 & 5.9e-1 & 4.6 \\
SETTLS         & 1.2e-4 & 6.0e-3 & 5.9e-1 & 4.6 \\
endpoint       & 1.2e-4 & \textbf{5.8e-3} & \textbf{5.9e-1} & \textbf{4.4} \\ \hline
\textbf{Disspiative GHD \eqref{eq:GHD_equation_nonlin}: Diffusion} & & & & \\ \hline
Crank-Nicolson & 9.4e-5 & 1.8e-4 & 9.6e-1 & 8.1 \\
midpoint       & 2.5e-4 & 3.4e-4 & 9.6e-1 & \textbf{4.4} \\
SETTLS         & \textbf{6.5e-5} & 1.9e-4 & \textbf{9.6e-1} & \textbf{4.4} \\
endpoint       & 8.9e-5 & \textbf{6.3e-5} & 9.6e-1 & 4.6 \\
\hline
\end{tabular}
\caption{\label{tab:summary}
Summary of the quantum Newton's cradle benchmarks. For each solver, the relative error of $N$, $E$, and $S$ integrated over the final oscillation of the cradle is reported. Each measure is computed for the benchmark featuring the highest time resolution. Additionally, the computationally runtime of each scheme normalized to the runtime of the RK1 (explicit) scheme is listed. For each metric, the best performing solver is highlighted in bold.
}
\end{table}

A summary of the benchmarks of the various schemes on the quantum Newton's cradle problem can be seen in table~\ref{tab:summary}.
Evaluating the accuracy of the numerical solutions is not straightforward, as no exact solution to the problem exists and the difficulty of solving the various equations, in particular the dressing equation, depends on the local state $\vartheta(z, t)$.
Therefore, we quantify the accuracy of the schemes by integrating the relative error of the particle number, total energy, and entropy over the final (tenth) oscillation period of the cradle.
Note, the measure is only computed for the benchmarks with the highest time resolution, that is $N_{\mathrm{steps}} = 2000$ for solving the purely advective GHD equation \eqref{eq:GHD_equation}
and $N_{\mathrm{steps}} = 3000$ for the dissipative equation \eqref{eq:GHD_equation_nonlin}.
For solving the advective equation, we find that the fourth order Runge--Kutta schemes, in particular the explicit variant, result in the lowest integrated error, i.e. the most accurate solution.
Meanwhile, for solving the dissipative equations, all the implemented schemes exhibit similar levels of accuracy.

In the context of GHD, the by far most numerically expensive operation of performing a BSL time step is solving the dressing equation.
The explicit RK1 scheme is by far the simplest of the schemes implemented, as each time step only involves calculating the effective velocity (requiring the dressing equation to be solved twice), the calculation of the departure points via equation~\eqref{eq:RK_scheme}, and the final interpolation of the filling function.
Among these operations, calculating the effective velocity far eclipses the computational runtime of the others.
Therefore, we report the computationally runtime of each scheme in table~\ref{tab:summary} in units of the runtime of the explicit RK1 scheme.
As stated earlier, higher order schemes employing the exact derivatives of the velocity fields require solving the dressing equation an additional two times.
Hence, for solving the purely advective BSL equation, we find the computational runtime of said schemes to be around twice that of the explicit RK1 scheme.
Interestingly, the implicit and explicit schemes exhibit very similar runtimes; in each iteration of the implicit schemes, the velocity fields must be interpolated to new points in the phase-space, however, this is a relatively cheap operation compared to evaluating the field to begin with. 
Furthermore, going beyond second order in numerical scheme does not substantially increase the runtime, suggesting that GHD calculations may benefit from employing even higher order schemes.
Importantly, for non-GHD applications, we would expect a notable difference in the runtime between implicit and explicit schemes and a notable increase in runtime as the scheme order increases.
Additionally, we find that the Adams--Moulton schemes employing extrapolation from previous time points to obtain the velocity fields at time $t^{n+1}$ are very efficient, as the effective velocity must only be evaluated once per time step.
However, we again stress that the large discrepancy in runtime between the Adams--Moulton extrapolation schemes and the rest is specific to the GHD case.
Finally, for solving the dissipative GHD equations, we find that all schemes, besides the Crank-Nicolson scheme, exhibit similar runtimes.
The runtime of the latter is much higher, as both evaluating the collision integral and the diffusion kernel is computationally expensive and must be be done in each iteration of the scheme.

\section{Conclusion and outlook} \label{s:conclusion}

We  present dissipative Generalized HydroDynamic (GHD) equations as a kinetic nonlinear PDE proven to be very useful to model accurately the dynamics observed in quantum experiments for quasi one-dimensional systems of cold atoms.
Our main goal is to present, benchmark and discuss numerical methods for solving the GHD equations with a diffusion term or a Boltzmann type collission term.
In particular, we design novel backward semi-Lagrangian (BSL)  schemes of high-order, namely implicit/explicit
Runge--Kutta BSL schemes.
Despite a growing interest for the (dissipative) GHD equations, accompanied by an abundant literature on the subject, it is the first time that these equations are solved with such high-order numerical schemes. 
In order to prove the accuracy  of these new numerical methods, we perform some comparisons with other schemes in the literature. Numerical simulations of the example of "quantum Newton's cradle" setup show that implicit/explicit Runge--Kutta BSL schemes preserve a posteriori some integral invariants (number of quasi-particles, energy, entropies) of the system in an accurate and satisfactory manner. 
These observations prevail as well in the case of numerical methods including the dissipative terms as without them. 
Finally, our numerical simulations of the dissipative GHD equations are consistent with the thermalization processes observed in experimental quasi one-dimensional devices initialized with a quantum Newton’s cradle setup~\cite{PhysRevLett.126.090602}.

There are several future extensions of our work: 
First we can develop other high-order numerical schemes for solving the dressing operation, such as the continuous-Galerkin or discontinuous-Galerkin methods, but  also some spectral (Fourier-like) methods to obtain a faster dressing transformation. 
Second we can devise other higher-order in time BSL-hybrid schemes for solving the dissipative GHD equations by using the strategy of exponential integrators \cite{BKV98, CM02, CK09, HO10, Tok11, CKV16, GP16, LO16, PS19}. 
We can also design higher-order schemes for the advection (Vlasov) operator by considering discontinuous-Galerkin schemes \cite{CS01, HGMM12} or, even better, semi-Lagrangian discontinuous-Galerkin methods \cite{RS11, QS11, BDM17}. 
In addition, we can consider other high-order and fast methods such as spectral methods (Fourier--Galerkin,  discontinuous-Galerkin) or real methods (discrete-velocity) for solving the Boltzmann-like collision and diffusion operators \cite{DP14, MPR13, JAH16, GHHH17, PR21}. Finally, it would be interesting to prove a $H$-theorem for the Boltzmann-type collision integral introduced in Section~\ref{sss:Colop}.

\appendix
\section*{Appendix}
\section{Some properties of the GHD equations}
\label{Appendix:GHD}
In this appendix, we show some properties of the GHD equations written either in conservative form through
equations \eqref{eq:GHD_equation_conservative}-\eqref{eq:aeff_rhop} or in advective form through
equations \eqref{eq:GHD_equation}, \eqref{eq:effective_velocity} and \eqref{eq:effective_acceleration}.

\subsection{Equation for the density of states $\rho_{\mathrm{s}}$}
Here, we show that  the density of states $\rho_{\mathrm{s}}$, defined by \eqref{def:rhos}, satisfies the same equation
as  $\rho_{\mathrm{p}}$, i.e.,
\begin{equation}
  \partial_{t} \rho_\mathrm{s}+ \partial_{z} \big( v^{\mathrm{eff}} \,
  \rho_\mathrm{s} \big) + \partial_{\theta} \big( a^{\mathrm{eff}} \, \rho_\mathrm{s} \big) =0.
  \label{eq:CLrhos_1}
\end{equation}
Using definition \eqref{def:rhos} for $\rho_{\mathrm{s}}$
and definition \eqref{eq:dressing} for the dressing operation, the density of states $\rho_{\mathrm{s}}$ rewrites as
\begin{equation} 
  \rho_{\mathrm{s}}(\theta ) =\frac{1}{2\pi} + \int_\R \mathrm{d}\theta'\, 
  T (\theta -\theta')\rho_{\mathrm{p}}(\theta') = \frac{1}{2\pi} + T\ast \rho_{\mathrm{p}},
\label{u:2b} 
\end{equation}
where the symbol $\ast$ denotes the standard convolution in the rapidity variable $\theta$.
Using \eqref{u:2b} equations \eqref{eq:veff_rhop}-\eqref{eq:aeff_rhop} rewrites as
\begin{eqnarray}
 \label{u:11}
&&   \rho_{\mathrm{s}}v^{\mathrm{eff}}=
   \frac{\partial_\theta \epsilon}{2\pi} + T \ast(v^{\mathrm{eff}} \rho_{\mathrm{p}}), \\
&& \rho_{\mathrm{s}}a^{\mathrm{eff}}=
   -\frac{\partial_z \epsilon}{2\pi} + T \ast(a^{\mathrm{eff}} \rho_{\mathrm{p}}).
 \label{u:12}
\end{eqnarray}
Using \eqref{u:2b}-\eqref{u:12} and the properties of the convolution operation, we obtain
from equation \eqref{eq:GHD_equation_conservative} for the density of quasi-particle $\rho_{\mathrm{p}}$,
\begin{eqnarray*}
  0&=&T \ast [
  \partial_t \rho_{\mathrm{p}}+\partial_z (v^{\mathrm{eff}}\rho_{\mathrm{p}})+\partial_\theta (a^{\mathrm{eff}} \rho_{\mathrm{p}})]\\
  &=&\partial_t(T \ast \rho_{\mathrm{p}}) + \partial_z (T \ast[v^{\mathrm{eff}}\rho_{\mathrm{p}}])
  + \partial_\theta(T \ast[a^{\mathrm{eff}} \rho_{\mathrm{p}}])\\
  &=& \partial_t(\rho_{\mathrm{s}}) + \partial_z(\rho_{\mathrm{s}} v^{\mathrm{eff}} -\partial_\theta \epsilon/(2\pi))
  +  \partial_\theta(\rho_{\mathrm{s}} a^{\mathrm{eff}} +\partial_z \epsilon/(2\pi))\\
  &=&\partial_t(\rho_{\mathrm{s}}) + \partial_z(\rho_{\mathrm{s}} v^{\mathrm{eff}})
  +  \partial_\theta(\rho_{\mathrm{s}} a^{\mathrm{eff}}),
\end{eqnarray*}
which is equation \eqref{eq:CLrhos_1}.

\subsection{Integral invariants}
\label{Appendix:GHD:Inv}
Here, we show that the GHD equations preserve in time the total number of quasi-particles $N$ defined
by \eqref{CL:N}, the total density of states $J$ defined by \eqref{CL:Gamma}, the total energy $E$ defined
by \eqref{CL:E}, and the infinite number of entropies $S$ defined by \eqref{CL:E}. We start with  $N$
and  $J$. Time invariance of $N$ and $J$ is obvious from  integration in the phase space $(z,\theta)$
of respectively the conservation law  \eqref{eq:GHD_equation_conservative} for $\rho_{\mathrm{p}}$ and 
the conservation law  \eqref{eq:CLrhos} for $\rho_{\mathrm{s}}$. For the time invariance of $E$, we first
observe that for any functions $\theta\mapsto g \in L^2(\R)$ and $\theta \mapsto h  \in L^2(\R) $, we have
\begin{equation}
  \langle (1-\widehat{T}f)g, h \rangle= \langle g, (1-f\widehat{T})h \rangle,
  \label{transRel}
\end{equation}  
where
\[
\langle g, h \rangle = \int_\R d\theta \, g(\theta) h(\theta),
\]
is the standard scalar product in $L^2(\R)$.
In other words, \eqref{transRel} means that $(1-\widehat{T}f)^T = (1-f\widehat{T})$, where the operator
$\widehat{K}^T$ is the transpose or adjoint operator of the operator $\widehat{K}$ induced by the scalar product
$\langle \cdot, \cdot\rangle$.
Indeed relation \eqref{transRel} is a consequence of the following property of the convolution operator,
$ \langle g\ast T,h \rangle= \langle g, T\ast h \rangle$, with $T$ an even function.
Using \eqref{transRel} and definition \eqref{eq:dressing_short} of the dressing operator, we obtain
\begin{equation}
  \langle f g^{\mathrm{dr}}, h \rangle= \langle g, f h^{\mathrm{dr}} \rangle.
  \label{transdr}
\end{equation}
Indeed, using \eqref{transRel} and  definition \eqref{eq:dressing_short}, we obtain
\begin{eqnarray*}
  \langle f g^{\mathrm{dr}}, h \rangle &=&   \langle f (1-\widehat{T}f)^{-1}g, h \rangle 
  =  \langle  (1-\widehat{T}f)^{-1}g, fh \rangle 
  =  \langle g,  (1-\widehat{T}f)^{-T}fh \rangle 
  =\langle g, (1-f\widehat{T})^{-1}f h \rangle\\
  &=&\langle g, [f(1-\widehat{T}f)(1/f)]^{-1}f h \rangle
  =\langle g, f(1-\widehat{T}f)^{-1} h \rangle
  =  \langle g, f h^{\mathrm{dr}} \rangle.
\end{eqnarray*}
Now, using integration by parts, the relation $\rho_{\mathrm{p}}=\rho_{\mathrm{s}} f$, the identity $\rho_{\mathrm{s}}=1^{\mathrm{dr}}/(2\pi)$,
the definitions \eqref{eq:effective_velocity}-\eqref{eq:effective_acceleration}, and the property \eqref{transdr}, we
obtain from he conservation law  \eqref{eq:GHD_equation_conservative} for $\rho_{\mathrm{p}}$,
\begin{eqnarray*}
  \frac{\mathrm{d}}{\mathrm{d}t} E(t)&=& \frac{\mathrm{d}}{\mathrm{d}t}\Big(\int_\R \mathrm{d}z \int_\R \mathrm{d}\theta\, \rho_{\mathrm{p}} \epsilon \Big)\\
  &=&\int_\R \mathrm{d}z \int_\R \mathrm{d}\theta\, \partial_t\rho_{\mathrm{p}} \epsilon \\
  &=& -\int_\R \mathrm{d}z \int_\R \mathrm{d}\theta\, \{\partial_z(\rho_{\mathrm{p}} v^{\mathrm{eff}}) +\partial_\theta(\rho_{\mathrm{p}} a^{\mathrm{eff}})\}  \epsilon\\
  &=& \int_\R \mathrm{d}z \int_\R \mathrm{d}\theta\, \{\rho_{\mathrm{p}} v^{\mathrm{eff}}\partial_z\epsilon +\rho_{\mathrm{p}} a^{\mathrm{eff}} \partial_\theta\epsilon\} \\
  &=&\frac{1}{2\pi}\int_\R \mathrm{d}z \int_\R \mathrm{d}\theta\, f\{ 1^{\mathrm{dr}}v^{\mathrm{eff}}\partial_z\epsilon +
  1^{\mathrm{dr}} a^{\mathrm{eff}} \partial_\theta\epsilon\} \\
  &=&\frac{1}{2\pi}\int_\R \mathrm{d}z \int_\R \mathrm{d}\theta\, f\{ (\partial_\theta\epsilon)^{\mathrm{dr}}\partial_z\epsilon +
  (-\partial_z\epsilon)^{\mathrm{dr}}\partial_\theta\epsilon\} \\
  &=&0.
\end{eqnarray*}
For the time invariance of $S$, using an integration by parts, and equations \eqref{eq:GHD_equation} and \eqref{eq:GHD_equation_conservative},
we obtain
\begin{eqnarray*}
  \frac{\mathrm{d}}{\mathrm{d}t} S(t)&=& \frac{\mathrm{d}}{\mathrm{d}t}\Big(\int_\R \mathrm{d}z \int_\R \mathrm{d}\theta\, \rho_{\mathrm{p}} \Psi(f) \Big)\\
  &=& \int_\R \mathrm{d}z \int_\R \mathrm{d}\theta\,\{ \Psi(f)\partial_t\rho_{\mathrm{p}}  + \rho_{\mathrm{p}} \Psi^\prime(f) \partial_t f\\
  &=& -\int_\R \mathrm{d}z \int_\R \mathrm{d}\theta\, \{ \Psi(f) [\partial_{z} ( v^{\mathrm{eff}} \,
    \rho_\mathrm{p} ) + \partial_{\theta} ( a^{\mathrm{eff}} \, \rho_\mathrm{p})] + \rho_\mathrm{p}  \Psi^\prime(f)
       [v^{\mathrm{eff}}\partial_z f + a^{\mathrm{eff}}\partial_\theta f] \} \\
 &=& 0.      
\end{eqnarray*}

\section{Integral invariants for the dissipative GHD equations with the diffusion operator}
\label{appendix:IdGHDD}
In this appendix, we prove that the dissipative GHD equations, where the right-hand side is given by the diffusion operator viz. equations \eqref{eqn:V-DCF}-\eqref{eqn:DCF},  preserves the total number of quasi-particles $N$ defined
by \eqref{CL:N}, and the total energy $E$ defined by \eqref{CL:E}. Since in Appendix~\ref{Appendix:GHD:Inv}  we have already shown the conservation of the total number of quasi-particle and of the total energy for the transport part, i.e. for the left-hand side of equation \eqref{eqn:V-DCF}, it only remains to show that these quantities are preserved by the diffusion operator \eqref{eqn:DCF}. From the conservative form of the diffusion operator \eqref{eqn:DCF}, an integration of the latter in the phase space $(z,\theta)$ gives straightforwardly the conservation of the total number quasi-particles $N$. We now show the conservation of the total energy $E$ in the case
where the potential $W$, entering in the definition of the single-particle energy $\epsilon$ given by \eqref{def:hamiltonian}, is independent of the rapidity variable $\theta$, namely $W(z,\theta)=V(z)$. Taking the diffusion operator \eqref{eqn:DCF}, multiplying it by $\epsilon$, integrating the result in the phase space $(z,\theta)$ and using an integration by parts
in the position variable $z$, we obtain
\begin{equation}
\label{eq:ECdDGHD:1}
\int_\R \mathrm{d} z \int_\R \mathrm{d} \theta\,
\epsilon\mathcal{D}_{\rho_{\mathrm{p}}}[\rho_{\mathrm{p}}]
= \int_\R \mathrm{d} z \int_\R \mathrm{d} \theta\,
(-\partial_zV) (1-f\widehat{T})^{-1}\rho_{\mathrm{s}}\widehat{D}
\rho_{\mathrm{s}}^{-1}
(1-f\widehat{T})^{-1}\rho_{\mathrm{p}}.
\end{equation}
Using transposition \eqref{transRel}, definition \eqref{eq:effective_acceleration}, 
and since $g^{\mathrm{dr}}=(1-\widehat{T}f)^{-1}g$, for any function $g$,
we obtain from \eqref{eq:ECdDGHD:1},
\begin{eqnarray}
\label{eq:ECdDGHD:2}
\int_\R \mathrm{d} z \int_\R \mathrm{d} \theta\,
\epsilon\mathcal{D}_{\rho_{\mathrm{p}}}[\rho_{\mathrm{p}}]
&=&\int_\R \mathrm{d} z \int_\R \mathrm{d} \theta\,
\big[(1-\widehat{T}f)^{-1}(-\partial_zV)\big]\rho_{\mathrm{s}}\widehat{D}
\rho_{\mathrm{s}}^{-1}
(1-f\widehat{T})^{-1}\rho_{\mathrm{p}} \nonumber\\
&= & \int_\R \mathrm{d} z \int_\R \mathrm{d} \theta\,
(-\partial_z\epsilon)^{\mathrm{dr}}\rho_{\mathrm{s}}\widehat{D}
\rho_{\mathrm{s}}^{-1}
(1-f\widehat{T})^{-1}\rho_{\mathrm{p}} \nonumber\\
&=&\frac{1}{2\pi}\int_\R \mathrm{d} z \int_\R \mathrm{d} \theta\,
a^{\mathrm{eff}}\rho_{\mathrm{s}}^2\widehat{D}
\rho_{\mathrm{s}}^{-1}
(1-f\widehat{T})^{-1}\rho_{\mathrm{p}}.
\end{eqnarray}
Since the potential $V=V(z)$ depends only on the position variable $z$,
we have $a^{\mathrm{eff}}=-\partial_z V$. Using this and 
setting $g:=\rho_{\mathrm{s}}^{-1}
(1-f\widehat{T})^{-1}\rho_{\mathrm{p}}$, 
equation \eqref{eq:ECdDGHD:2} becomes
 \begin{equation}
 \label{eq:ECdDGHD:3}
\int_\R \mathrm{d} z \int_\R \mathrm{d} \theta\,
\epsilon\mathcal{D}_{\rho_{\mathrm{p}}}[\rho_{\mathrm{p}}]
= \frac{1}{2\pi}\int_\R \mathrm{d} z \, (-\partial_z V)\int_\R \mathrm{d} \theta\,
\rho_{\mathrm{s}}^2\widehat{D}g.
\end{equation}
Now, inserting the definition \eqref{def:kernel:D} for the kernel $D$ into
equation \eqref{eq:ECdDGHD:3}, we obtain
\begin{eqnarray*}
\int_\R \mathrm{d} z \int_\R \mathrm{d} \theta\,
\epsilon\mathcal{D}_{\rho_{\mathrm{p}}}[\rho_{\mathrm{p}}]
&=& \frac{1}{2\pi}\int_\R \mathrm{d} z \, (-\partial_z V)\int_\R \mathrm{d} \theta\, \int_\R \mathrm{d}\alpha\, g(\alpha) \Big[
\delta(\theta -\alpha)\int_\R \mathrm{d}\gamma\, W(\gamma,\alpha)
-W(\theta,\alpha)\Big]\\
&=&\frac{1}{2\pi}\int_\R \mathrm{d} z \, (-\partial_z V)
\bigg(\int_\R \mathrm{d}\gamma \int_\R \mathrm{d}\alpha\,
g(\alpha)W(\gamma,\alpha)-
\int_\R \mathrm{d}\theta \int_\R \mathrm{d}\alpha\,
g(\alpha)W(\theta,\alpha)
\bigg)=0,
\end{eqnarray*}
which ends the proof of the conservation of the total energy $E$. 

\section{Construction of Boltzmann collision integral for transverse state-changing collisions} \label{appendix:collision}

In this appendix, we outline the construction of the Boltzmann-type collision integral, originally originally featured in \cite{PhysRevLett.126.090602}.
Given a perturbation to an integrable Hamiltonian, which only respect a few of the conservation laws of the unperturbed Hamiltonian, an associated collision integral can be formally derived following perturbation theory~\cite{PhysRevB.101.180302, PhysRevLett.127.130601}.
The conservation laws of the perturbation (typically number of particles, momentum, and energy) are explicitly included in the derivation, often determining the rapidities involved in the scattering process.
However, for most experimentally relevant processes, such formal derivation is infeasible and one must construct the collision integral following a more phenomenological approach, as well shall do here.

Assuming the majority of atoms remains in the transverse ground state, we can restrict our treatment
to the three lowest states of the transverse potential, here denoted by the indices $n= 0,1,2$.
Within this effective three-level system, two excitation (and de-excitation) events are possible by virtue of parity:
(1) Two atoms in the ground state collide and both are excited to the first excited state, or (2) two atoms in the ground state collide and only one is excited to the second excited state.
Similar selection rules exist for de-exciting collisions.

Consider a state-changing collision of two quasi-particles with rapidities $\theta$ and $\theta'$, and let $\theta_{\pm}$ and $\theta_{\pm}^\prime$ be their rapidities following the collisions.
Here, the subscript "$-$" denotes an exciting collision, while "$+$" denotes a de-exciting one.
Following conservation of energy we must have
\begin{equation*}
    \theta^2 + (\theta^\prime)^2 = \theta_{\pm}^2 + (\theta_{\pm}^\prime)^2 \mp 4 l_{\perp}^{-2} + \Delta E (\theta, \theta^\prime , \theta_\pm, \theta_{\pm}^\prime),
\end{equation*}
The term $\mp 4 l_{\perp}^{-2}$ denotes the change in transverse potential energy following the state-changing collision.
Meanwhile, the conservation of momentum implies
\begin{equation*}
  \theta + \theta^\prime = \theta_{\pm} + \theta_{\pm}^\prime + \Delta P (\theta, \theta^\prime , \theta_\pm, \theta_{\pm}^\prime).
\end{equation*}
The terms $\Delta E$ and $\Delta P$ describe a shift in the total energy and momentum, respectively, resulting from a reconfiguration of all local rapidities following the collision.
This effect is captured by the so-called backflow function, and is a result of the 1D system being intrinsically strongly correlated~\cite{korepin_bogoliubov_izergin_1993}.
Finding the post-collision rapidities in the presence of backflow is rather difficult, as the backflow itself is a functional of the current state of the system (characterized by $\rho_{\mathrm{p}}$ or $f$).
Therefore, to construct a numerically tractable model, the effect of the backflow was ignored in \cite{PhysRevLett.126.090602}.
Note that for most experimentally relevant systems this is a reasonable approximation. 
Neglecting the backflow, the post-collision rapidities becomes independent of the state and read
$\theta_\pm = \frac 12 (\theta +\theta^\prime )+\frac 12 (\theta -\theta^\prime )\sqrt{1\pm 8/[(\theta -\theta^\prime )l_\perp ]^2}$ 
and
$\theta_\pm^\prime = \frac 12 (\theta +\theta^\prime )-\frac 12 (\theta -\theta^\prime )\sqrt{1\pm 8/[(\theta -\theta^\prime )l_\perp ]^2}$.
Simplifying the scattering event to that of just two atoms colliding in a waveguide, the resulting scattering cross section reads 
\begin{equation*} 
    P_\updownarrow (\theta_1,\, \theta_2)= \frac{4 c^2\theta_1\theta_2}{\theta_1^2\theta_2^2 +c^2(\theta_1+\theta_2 )^2} \; .
\end{equation*} 
Following these considerations, the Boltzmann collision integral can be constructed.
Within GHD, Boltzmann collision integrals are typically constructed in the conservative form in order to explicitly assure the conservation of particle number.
For the transverse state-changing collisions, each possible excitation (and de-excitation) event has its own corresponding collision integral $\mathcal{I}_{\mathrm{p}, \mathrm{h}} ^\pm [\rho_{\mathrm{p}}]$.
Each of these collision integrals describes the destruction of two quasi-particles (i.e. the creation of two holes) at the incoming rapidities and creation of two quasi-particles at the outgoing ones, thus conserving the total number of particles.
The expressions for these collisions integrals can be found in Section~\ref{sss:Colop}.
Although a number of simplifications were made in order to arrive at the collision integral, it has been demonstrated to successfully describe experimental observations~\cite{PhysRevLett.126.090602}.

 \section*{Acknowledgement}
We acknowledge support from the Austrian Science Fund (FWF) via the grants SFB F65 "Taming Complexity in PDE systems" and, together with the German Research Foundation (DFG), via the Research Unit FOR 2724 “Thermal machines in the thermal world.” The work was further supported by the FQXI program on “Informations as fuel” and the ESQ Discovery Grant “Emergence of physical laws: from mathematical foundations to applications in many body physics” of the Austrian Academy of Sciences (\"{O}AW), and by the Vienna Science and Technology Fund (WWTF) project MA16-066 "SEQUEX".\\ 
N.B. acknowledges the hospitality of the Wolfgang Pauli Institute Vienna which boosted this research.

\bibliography{references}

\begin{thebibliography}{100}

\bibitem{ACIOLI199775}
P.~H. Acioli.
\newblock Review of quantum {M}onte {C}arlo methods and their applications.
\newblock {\em J. Mol. Struct.: THEOCHEM}, 394(2):75--85, 1997.
\newblock Proceedings of the Eighth Brazilian Symposium of Theoretical
  Chemistry.

\bibitem{Alba_2021}
V.~Alba, B.~Bertini, M.~Fagotti, L.~Piroli, and P.~Ruggiero.
\newblock Generalized-hydrodynamic approach to inhomogeneous quenches:
  correlations, entanglement and quantum effects.
\newblock {\em J. Stat. Mech. Theory Exp.}, 2021(11):114004, nov 2021.

\bibitem{PhysRevLett.123.130602}
A.~Bastianello, V.~Alba, and J.-S. Caux.
\newblock {G}eneralized {H}ydrodynamics with space-time inhomogeneous
  interactions.
\newblock {\em Phys. Rev. Lett.}, 123:130602, Sep 2019.

\bibitem{Bastianello_2022}
A.~Bastianello, B.~Bertini, B.~Doyon, and R.~Vasseur.
\newblock Introduction to the special issue on emergent hydrodynamics in
  integrable many-body systems.
\newblock {\em J. Stat. Mech. Theory Exp.}, 2022(1):014001, jan 2022.

\bibitem{PhysRevLett.125.240604}
A.~Bastianello, A.~De~Luca, B.~Doyon, and J.~De~Nardis.
\newblock Thermalization of a trapped one-dimensional {B}ose gas via diffusion.
\newblock {\em Phys. Rev. Lett.}, 125:240604, Dec 2020.

\bibitem{Bastianello_2021}
A.~Bastianello, A.~D. Luca, and R.~Vasseur.
\newblock Hydrodynamics of weak integrability breaking.
\newblock {\em J. Stat. Mech. Theory Exp.}, 2021(11):114003, nov 2021.

\bibitem{bertini2016transport}
B.~Bertini, M.~Collura, J.~De~Nardis, and M.~Fagotti.
\newblock Transport in out-of-equilibrium {XXZ} chains: Exact profiles of
  charges and currents.
\newblock {\em \textit{Phys. Rev. Lett.}}, 117(20):207201, 2016.

\bibitem{Bes04}
N.~Besse.
\newblock Convergence of a semi-{L}agrangian scheme for the one-dimensional
  {V}lasov--{P}oisson system.
\newblock {\em SIAM J. Numer. Anal.}, 42:350--382, 2004.

\bibitem{Bes08}
N.~Besse.
\newblock Convergence of a high-order semi-{L}agrangian scheme with propagation
  of gradients for the {V}lasov--{P}oisson system.
\newblock {\em SIAM J. Numer. Anal.}, 46:639--670, 2008.

\bibitem{BDM17}
N.~Besse, E.~Deriaz, and E.~Madaule.
\newblock Adaptive multiresolution semi-{L}agrangian discontinuous {G}alerkin
  methods for the {V}lasov equations.
\newblock {\em J. Comput. Phys.}, 332, 2017.

\bibitem{BM08}
N.~Besse and M.~Mehrenberger.
\newblock Convergence of classes of high-order semi-{L}agrangian schemes for
  the {V}lasov--{P}oisson system.
\newblock {\em Math. Comp.}, 77:93--123, 2008.

\bibitem{BesseSonnenJCP03}
N.~Besse and E.~Sonnendr\"ucker.
\newblock Semi-{L}agrangian schemes for the {V}lasov equation on an
  unstructured mesh of phase space.
\newblock {\em J. Comput. Phys.}, 191:341--376, 2003.

\bibitem{Bethe1931}
H.~Bethe.
\newblock Zur {T}heorie der {M}etalle.
\newblock {\em Zeitschrift f{\"u}r Physik}, 71(3):205--226, Mar 1931.

\bibitem{BKV98}
G.~Beylkin, J.~M. Keiser, and L.~Vozovoi.
\newblock A new class of time discretization schemes for the solution of
  nonlinear {PDE}s.
\newblock {\em J. Comput. Phys.}, 147:362--387, 1998.

\bibitem{Bloch2012}
I.~Bloch, J.~Dalibard, and S.~Nascimb{\`e}ne.
\newblock Quantum simulations with ultracold quantum gases.
\newblock {\em Nature Physics}, 8(4):267--276, Apr 2012.

\bibitem{RevModPhys.80.885}
I.~Bloch, J.~Dalibard, and W.~Zwerger.
\newblock Many-body physics with ultracold gases.
\newblock {\em Rev. Mod. Phys.}, 80:885--964, Jul 2008.

\bibitem{Borsi_2021}
M.~Borsi, B.~Pozsgay, and L.~Pristyák.
\newblock Current operators in integrable models: a review.
\newblock {\em J. Stat. Mech. Theory Exp.}, 2021(9):094001, sep 2021.

\bibitem{Bouchoule_2022}
I.~Bouchoule and J.~Dubail.
\newblock {G}eneralized {H}ydrodynamics in the one-dimensional {B}ose gas:
  theory and experiments.
\newblock {\em J. Stat. Mech. Theory Exp.}, 2022(1):014003, jan 2022.

\bibitem{Bulchandani_2021}
V.~B. Bulchandani, S.~Gopalakrishnan, and E.~Ilievski.
\newblock Superdiffusion in spin chains.
\newblock {\em J. Stat. Mech. Theory Exp.}, 2021(8):084001, aug 2021.

\bibitem{PhysRevLett.119.220604}
V.~B. Bulchandani, R.~Vasseur, C.~Karrasch, and J.~E. Moore.
\newblock Solvable hydrodynamics of quantum integrable systems.
\newblock {\em Phys. Rev. Lett.}, 119:220604, Nov 2017.

\bibitem{PhysRevB.97.045407}
V.~B. Bulchandani, R.~Vasseur, C.~Karrasch, and J.~E. Moore.
\newblock Bethe-{B}oltzmann hydrodynamics and spin transport in the {XXZ}
  chain.
\newblock {\em Phys. Rev. B}, 97:045407, Jan 2018.

\bibitem{Buca_2021}
B.~Buča, K.~Klobas, and T.~Prosen.
\newblock Rule 54: exactly solvable model of nonequilibrium statistical
  mechanics.
\newblock {\em J. Stat. Mech. Theory Exp.}, 2021(7):074001, jul 2021.

\bibitem{Calabrese_2016}
P.~Calabrese, F.~H.~L. Essler, and G.~Mussardo.
\newblock Introduction to ‘quantum integrability in out of equilibrium
  systems’.
\newblock {\em J. Stat. Mech. Theory Exp.}, 2016(6):064001, jun 2016.

\bibitem{castro2016emergent}
O.~A. Castro-Alvaredo, B.~Doyon, and T.~Yoshimura.
\newblock Emergent hydrodynamics in integrable quantum systems out of
  equilibrium.
\newblock {\em \textit{Phys. Rev. X}}, 6(4):041065, 2016.

\bibitem{PhysRevX.12.041032}
F.~Cataldini, F.~M\o{}ller, M.~Tajik, J.~a. Sabino, S.-C. Ji, I.~Mazets,
  T.~Schweigler, B.~Rauer, and J.~Schmiedmayer.
\newblock Emergent pauli blocking in a weakly interacting {B}ose gas.
\newblock {\em Phys. Rev. X}, 12:041032, Dec 2022.

\bibitem{10.21468/SciPostPhys.6.6.070}
J.-S. Caux, B.~Doyon, J.~Dubail, R.~Konik, and T.~Yoshimura.
\newblock {Hydrodynamics of the interacting {B}ose gas in the Quantum Newton
  Cradle setup}.
\newblock {\em SciPost Phys.}, 6:70, 2019.

\bibitem{PhysRevLett.110.257203}
J.-S. Caux and F.~H.~L. Essler.
\newblock Time evolution of local observables after quenching to an integrable
  model.
\newblock {\em Phys. Rev. Lett.}, 110:257203, Jun 2013.

\bibitem{RevModPhys.83.1405}
M.~A. Cazalilla, R.~Citro, T.~Giamarchi, E.~Orignac, and M.~Rigol.
\newblock One dimensional bosons: From condensed matter systems to ultracold
  gases.
\newblock {\em Rev. Mod. Phys.}, 83:1405--1466, Dec 2011.

\bibitem{CK09}
E.~Celledoni and B.~K. Kometa.
\newblock Semi-{L}agrangian {R}unge--{K}utta exponential integrators for
  convection dominated problems.
\newblock {\em J. Sci. Comput.}, 41:139--164, 2009.

\bibitem{CKV16}
E.~Celledoni, B.~K. Kometa, and O.~Verdier.
\newblock High-order semi-{L}agrangian methods for the incompressible
  {N}avier--{S}tokes equations.
\newblock {\em J. Sci. Comput.}, 66:91--115, 2016.

\bibitem{CS01}
B.~Cockburn and C.-W. Shu.
\newblock {R}unge--{K}utta discontinuous {G}alerkin methods for
  convection-dominated problems.
\newblock {\em J. Sci. Comput.}, 16, 2001.

\bibitem{CM02}
S.~M. Cox and P.~C. Matthews.
\newblock Exponential time differencing for stiff systems.
\newblock {\em J. Comput. Phys.}, 176:430--455, 2011.

\bibitem{Cubero_2021}
A.~C. Cubero, T.~Yoshimura, and H.~Spohn.
\newblock Form factors and {G}eneralized {H}ydrodynamics for integrable
  systems.
\newblock {\em J. Stat. Mech. Theory Exp.}, 2021(11):114002, nov 2021.

\bibitem{Daley_2004}
A.~J. Daley, C.~Kollath, U.~Schollwöck, and G.~Vidal.
\newblock Time-dependent density-matrix renormalization-group using adaptive
  effective hilbert spaces.
\newblock {\em J. Stat. Mech. Theory Exp.}, 2004(04):P04005, apr 2004.

\bibitem{DeB01}
C.~De~Boor.
\newblock {\em A practical guide to splines}, volume~27 of {\em Applied
  mathematical sciences}.
\newblock Springer, 2001.

\bibitem{PhysRevLett.121.160603}
J.~De~Nardis, D.~Bernard, and B.~Doyon.
\newblock Hydrodynamic diffusion in integrable systems.
\newblock {\em Phys. Rev. Lett.}, 121:160603, Oct 2018.

\bibitem{NBD19}
J.~De~Nardis, D.~Bernard, and B.~Doyon.
\newblock Diffusion in {G}eneralized {H}ydrodynamics and quasiparticle
  scattering.
\newblock {\em SciPost Phys.}, 6:1--72, 2019.

\bibitem{DP14}
G.~Dimarco and L.~Pareschi.
\newblock Numerical methods for kinetic equations.
\newblock {\em Acta Numer.}, 23:369--520, 2014.

\bibitem{10.21468/SciPostPhysLectNotes.18}
B.~Doyon.
\newblock {Lecture notes on Generalised Hydrodynamics}.
\newblock {\em SciPost Phys. Lect. Notes}, page~18, 2020.

\bibitem{PhysRevLett.119.195301}
B.~Doyon, J.~Dubail, R.~Konik, and T.~Yoshimura.
\newblock Large-scale description of interacting one-dimensional {B}ose gases:
  {G}eneralized {H}ydrodynamics supersedes conventional hydrodynamics.
\newblock {\em Phys. Rev. Lett.}, 119:195301, Nov 2017.

\bibitem{SciPostPhys.2.2.014}
B.~Doyon and T.~Yoshimura.
\newblock {A note on {G}eneralized {H}ydrodynamics: inhomogeneous fields and
  other concepts}.
\newblock {\em SciPost Phys.}, 2:014, 2017.

\bibitem{PhysRevLett.120.045301}
B.~Doyon, T.~Yoshimura, and J.-S. Caux.
\newblock Soliton gases and {G}eneralized {H}ydrodynamics.
\newblock {\em Phys. Rev. Lett.}, 120:045301, Jan 2018.

\bibitem{PhysRevLett.127.130601}
J.~Durnin, M.~J. Bhaseen, and B.~Doyon.
\newblock Nonequilibrium dynamics and weakly broken integrability.
\newblock {\em Phys. Rev. Lett.}, 127:130601, Sep 2021.

\bibitem{alma9918447213902959}
D.~R. Durran.
\newblock {\em Numerical methods for wave equations in geophysical fluid
  dynamics}.
\newblock Texts in applied mathematics 32. Springer, New York, 1999.

\bibitem{El_2021}
G.~A. El.
\newblock Soliton gas in integrable dispersive hydrodynamics.
\newblock {\em J. Stat. Mech. Theory Exp.}, 2021(11):114001, nov 2021.

\bibitem{FILBET2016171}
F.~Filbet and C.~Prouveur.
\newblock High order time discretization for backward semi-lagrangian methods.
\newblock {\em J. Comput. Appl. Math.}, 303:171--188, 2016.

\bibitem{RevModPhys.73.33}
W.~M.~C. Foulkes, L.~Mitas, R.~J. Needs, and G.~Rajagopal.
\newblock Quantum {M}onte {C}arlo simulations of solids.
\newblock {\em Rev. Mod. Phys.}, 73:33--83, Jan 2001.

\bibitem{PhysRevB.101.180302}
A.~J. Friedman, S.~Gopalakrishnan, and R.~Vasseur.
\newblock Diffusive hydrodynamics from integrability breaking.
\newblock {\em Phys. Rev. B}, 101:180302, May 2020.

\bibitem{GHHH17}
I.~M. Gamba, J.~R. Hauck, C.~D. Hauck, and J.~Hu.
\newblock A fast spectral method for the {B}oltzmann collision operator with
  general collision kernels.
\newblock {\em SIAM J. Sci. Comput.}, 39:B658--B676, 2017.

\bibitem{GAUDIN196755}
M.~Gaudin.
\newblock Un systeme a une dimension de fermions en interaction.
\newblock {\em Physics Letters A}, 24(1):55--56, 1967.

\bibitem{gaudin2014bethe}
M.~Gaudin and J.~Caux.
\newblock {\em The Bethe Wavefunction}.
\newblock Cambridge University Press, 2014.

\bibitem{GP16}
S.~Gaudreault and J.~A. Pudykiewicz.
\newblock An efficient exponential time integration method for the numerical
  solution of the shallow water equations on the sphere.
\newblock {\em J. Comput. Phys.}, 322:827--848, 2016.

\bibitem{RevModPhys.80.1215}
S.~Giorgini, L.~P. Pitaevskii, and S.~Stringari.
\newblock Theory of ultracold atomic fermi gases.
\newblock {\em Rev. Mod. Phys.}, 80:1215--1274, Oct 2008.

\bibitem{PhysRevLett.87.130402}
A.~G\"orlitz, J.~M. Vogels, A.~E. Leanhardt, C.~Raman, T.~L. Gustavson, J.~R.
  Abo-Shaeer, A.~P. Chikkatur, S.~Gupta, S.~Inouye, T.~Rosenband, and
  W.~Ketterle.
\newblock Realization of {B}ose-{E}instein condensates in lower dimensions.
\newblock {\em \textit{Phys. Rev. Lett.}}, 87(13):130402, Sep 2001.

\bibitem{PhysRevLett.87.160405}
M.~Greiner, I.~Bloch, O.~Mandel, T.~W. H\"ansch, and T.~Esslinger.
\newblock Exploring phase coherence in a 2{D} lattice of {B}ose-{E}instein
  condensates.
\newblock {\em \textit{Phys. Rev. Lett.}}, 87(16):160405, Oct 2001.

\bibitem{doi:10.1126/science.aal3837}
C.~Gross and I.~Bloch.
\newblock Quantum simulations with ultracold atoms in optical lattices.
\newblock {\em Science}, 357(6355):995--1001, 2017.

\bibitem{10.2307/2002064}
P.~C. Hammer and J.~W. Hollingsworth.
\newblock Trapezoidal methods of approximating solutions of differential
  equations.
\newblock {\em Mathematical Tables and Other Aids to Computation},
  9(51):92--96, 1955.

\bibitem{HGMM12}
E.~Heath, I.~M. Gamba, P.~J. Morrison, and C.~Michler.
\newblock A discontinuous {G}alerkin method for the {V}lasov--{P}oisson system.
\newblock {\em J. Comput. Phys.}, 231, 2012.

\bibitem{HO10}
M.~Hochbruck and A.~Ostermann.
\newblock Exponential integrators.
\newblock {\em Acta Numer.}, 19:209--286, 2010.

\bibitem{Hor02}
M.~Hortal.
\newblock The development and testing of a new two-time-level semi-{L}agrangian
  scheme ({SETTLS}) in the {ECMWF} forecast model.
\newblock {\em Q. J. R. Meteorol. Soc.}, 128:1671--1687, 2002.

\bibitem{JAH16}
S.~Jaiswal, A.~A. Alexeenko, and J.~Hu.
\newblock A discontinuous {G}alerkin fast spectral method for the full
  {B}oltzmann equation with general collision kernels.
\newblock {\em J. Comput. Phys.}, 378:178--208, 2019.

\bibitem{doi:10.1126/science.1100700}
T.~Kinoshita, T.~Wenger, and D.~S. Weiss.
\newblock Observation of a one-dimensional {T}onks-{G}irardeau gas.
\newblock {\em Science}, 305(5687):1125--1128, 2004.

\bibitem{kinoshita2006quantum}
T.~Kinoshita, T.~Wenger, and D.~S. Weiss.
\newblock A quantum {N}ewton's cradle.
\newblock {\em \textit{Nature}}, 440(7086):900, 2006.

\bibitem{korepin_bogoliubov_izergin_1993}
V.~E. Korepin, N.~M. Bogoliubov, and A.~G. Izergin.
\newblock {\em Quantum Inverse Scattering Method and Correlation Functions}.
\newblock Cambridge Monographs on Mathematical Physics. Cambridge University
  Press, 1993.

\bibitem{PhysRevLett.105.265302}
P.~Kr\"uger, S.~Hofferberth, I.~E. Mazets, I.~Lesanovsky, and J.~Schmiedmayer.
\newblock Weakly interacting {B}ose gas in the one-dimensional limit.
\newblock {\em Phys. Rev. Lett.}, 105(26):265302, Dec 2010.

\bibitem{Landau1933}
L.~D. Landau.
\newblock {\"Uber Die Bewegung der Elektronen in Kristallgitter}.
\newblock {\em Phys. Z. Sowjetunion}, 3:644--645, 1933.

\bibitem{doi:10.1126/science.1257026}
T.~Langen, S.~Erne, R.~Geiger, B.~Rauer, T.~Schweigler, M.~Kuhnert,
  W.~Rohringer, I.~E. Mazets, T.~Gasenzer, and J.~Schmiedmayer.
\newblock Experimental observation of a generalized {G}ibbs ensemble.
\newblock {\em Science}, 348(6231):207--211, 2015.

\bibitem{RevModPhys.78.17}
P.~A. Lee, N.~Nagaosa, and X.-G. Wen.
\newblock Doping a mott insulator: Physics of high-temperature
  superconductivity.
\newblock {\em Rev. Mod. Phys.}, 78:17--85, Jan 2006.

\bibitem{10.21468/SciPostPhys.9.4.058}
C.~Li, T.~Zhou, I.~Mazets, H.-P. Stimming, F.~S. Møller, Z.~Zhu, Y.~Zhai,
  W.~Xiong, X.~Zhou, X.~Chen, and J.~Schmiedmayer.
\newblock {Relaxation of Bosons in One Dimension and the Onset of Dimensional
  Crossover}.
\newblock {\em SciPost Phys.}, 9:58, 2020.

\bibitem{lieb2013mathematical}
E.~Lieb and D.~Mattis.
\newblock {\em Mathematical Physics in One Dimension: Exactly Soluble Models of
  Interacting Particles}.
\newblock Perspectives in physics. Elsevier Science, 2013.

\bibitem{LL2}
E.~H. Lieb.
\newblock Exact analysis of an interacting {B}ose gas. {II}. the excitation
  spectrum.
\newblock {\em Phys. Rev.}, 130:1616--1624, May 1963.

\bibitem{LL1}
E.~H. Lieb and W.~Liniger.
\newblock Exact analysis of an interacting {B}ose gas. {I}. the general
  solution and the ground state.
\newblock {\em Phys. Rev.}, 130:1605--1616, May 1963.

\bibitem{PhysRevB.103.L060302}
J.~Lopez-Piqueres, B.~Ware, S.~Gopalakrishnan, and R.~Vasseur.
\newblock Hydrodynamics of nonintegrable systems from a relaxation-time
  approximation.
\newblock {\em Phys. Rev. B}, 103:L060302, Feb 2021.

\bibitem{LO16}
V.~T. Luan and A.~Ostermann.
\newblock Parallel exponential {R}osenbrock methods.
\newblock {\em Comput. Math. Appl.}, 71:1137--1150, 2016.

\bibitem{malvania2020generalized}
N.~Malvania, Y.~Zhang, Y.~Le, J.~Dubail, M.~Rigol, and D.~S. Weiss.
\newblock {G}eneralized {H}ydrodynamics in strongly interacting 1d {B}ose
  gases.
\newblock {\em Science}, 373(6559):1129--1133, 2021.

\bibitem{Mannella2005}
N.~Mannella, W.~L. Yang, X.~J. Zhou, H.~Zheng, J.~F. Mitchell, J.~Zaanen, T.~P.
  Devereaux, N.~Nagaosa, Z.~Hussain, and Z.-X. Shen.
\newblock Nodal quasiparticle in pseudogapped colossal magnetoresistive
  manganites.
\newblock {\em Nature}, 438(7067):474--478, Nov 2005.

\bibitem{MATLAB:2020}
MATLAB.
\newblock {\em version 9.8.0 (R2020a)}.
\newblock The MathWorks Inc., Natick, Massachusetts, 2020.

\bibitem{https://doi.org/10.48550/arxiv.2205.15871}
F.~M{\o}ller, S.~Erne, N.~J. Mauser, J.~Schmiedmayer, and I.~E. Mazets.
\newblock Bridging effective field theories and {G}eneralized {H}ydrodynamics.
\newblock {\em arXiv:2205.15871}, 2022.

\bibitem{PhysRevLett.126.090602}
F.~M\o{}ller, C.~Li, I.~Mazets, H.-P. Stimming, T.~Zhou, Z.~Zhu, X.~Chen, and
  J.~Schmiedmayer.
\newblock Extension of the {G}eneralized {H}ydrodynamics to the dimensional
  crossover regime.
\newblock {\em Phys. Rev. Lett.}, 126:090602, Mar 2021.

\bibitem{10.21468/SciPostPhysCore.3.2.016}
F.~S. M{\o}ller, G.~Perfetto, B.~Doyon, and J.~Schmiedmayer.
\newblock {Euler-scale dynamical correlations in integrable systems with fluid
  motion}.
\newblock {\em SciPost Phys. Core}, 3:016, 2020.

\bibitem{10.21468/SciPostPhys.8.3.041}
F.~S. M\o{}ller and J.~Schmiedmayer.
\newblock {Introducing i{F}luid: a numerical framework for solving
  hydrodynamical equations in integrable models}.
\newblock {\em SciPost Phys.}, 8:41, 2020.

\bibitem{MPR13}
C.~Mouhot, L.~Pareschi, and T.~Rey.
\newblock Convolution decomposition and fast summation methods for
  discrete-velocity approximations of the {B}oltzmann equation.
\newblock {\em ESAIM Math. Model. Numer. Anal.}, 47:1513--1531, 2013.

\bibitem{DeNardis_2022}
J.~D. Nardis, B.~Doyon, M.~Medenjak, and M.~Panfil.
\newblock Correlation functions and transport coefficients in generalised
  hydrodynamics.
\newblock {\em J. Stat. Mech. Theory Exp.}, 2022(1):014002, jan 2022.

\bibitem{PhysRev.112.309}
R.~Orbach.
\newblock Linear antiferromagnetic chain with anisotropic coupling.
\newblock {\em Phys. Rev.}, 112:309--316, Oct 1958.

\bibitem{https://doi.org/10.48550/arxiv.2205.06492}
M.~Panfil, S.~Gopalakrishnan, and R.~M. Konik.
\newblock Thermalization of interacting quasi-one-dimensional systems.
\newblock 2022.

\bibitem{PR21}
L.~Pareschi and T.~Rey.
\newblock Moment preserving {F}ourier--{G}alerkin spectral methods and
  application to the {B}oltzmann equation.
\newblock {\em SIAM J. Num. Anal.}, 60, 2021.

\bibitem{PS19}
P.~S. Peixoto and M.~Schreiber.
\newblock Semi-{L}agrangian exponential integrator with application to the
  rotating shallow water equations.
\newblock {\em SIAM J. Sci. Comput.}, 41:B903--B928, 2019.

\bibitem{Pollet_2012}
L.~Pollet.
\newblock Recent developments in quantum {M}onte {C}arlo simulations with
  applications for cold gases.
\newblock {\em Rep. Progr. Phys.}, 75(9):094501, aug 2012.

\bibitem{QS11}
J.-M. Qiu and C.-W. Shu.
\newblock Positivity preserving semi-{L}agrangian discontinuous {G}alerkin
  formulation: theoretical analysis and application to the {V}lasov--{P}oisson
  system.
\newblock {\em J. Comput. Phys.}, 230, 2011.

\bibitem{PhysRevLett.103.100403}
M.~Rigol.
\newblock Breakdown of thermalization in finite one-dimensional systems.
\newblock {\em Phys. Rev. Lett.}, 103:100403, Sep 2009.

\bibitem{Rigol2008}
M.~Rigol, V.~Dunjko, and M.~Olshanii.
\newblock {Thermalization and its mechanism for generic isolated quantum
  systems}.
\newblock {\em Nature}, 452(7189):854--858, apr 2008.

\bibitem{PhysRevLett.98.050405}
M.~Rigol, V.~Dunjko, V.~Yurovsky, and M.~Olshanii.
\newblock Relaxation in a completely integrable many-body quantum system: An ab
  initio study of the dynamics of the highly excited states of 1d lattice
  hard-core bosons.
\newblock {\em Phys. Rev. Lett.}, 98:050405, Feb 2007.

\bibitem{RS11}
J.~A. Rossmanith and D.~C. Seal.
\newblock A positivity-preserving high-order semi-lagrangian discontinuous
  {G}alerkin scheme for the {V}lasov--{P}oisson equations.
\newblock {\em J. Comput. Phys.}, 230, 2011.

\bibitem{PhysRevLett.124.140603}
P.~Ruggiero, P.~Calabrese, B.~Doyon, and J.~Dubail.
\newblock Quantum {G}eneralized {H}ydrodynamics.
\newblock {\em Phys. Rev. Lett.}, 124:140603, Apr 2020.

\bibitem{schemmer2019generalized}
M.~Schemmer, I.~Bouchoule, B.~Doyon, and J.~Dubail.
\newblock Generalized {H}ydrodynamics on an atom chip.
\newblock {\em Phys. Rev. Lett.}, 122(9):090601, 2019.

\bibitem{RevModPhys.77.259}
U.~Schollw\"ock.
\newblock The density-matrix renormalization group.
\newblock {\em Rev. Mod. Phys.}, 77:259--315, Apr 2005.

\bibitem{SCHOLLWOCK201196}
U.~Schollwöck.
\newblock The density-matrix renormalization group in the age of matrix product
  states.
\newblock {\em Ann. Physics}, 326(1):96--192, 2011.
\newblock January 2011 Special Issue.

\bibitem{Sch07}
L.~L. Schumaker.
\newblock {\em Spline functions: basic theory}.
\newblock Cambridge Mathematical Library. Cambridge University Press, 2007.

\bibitem{SP92}
P.~Smolarkiewicz and J.~A. Pudykiewicz.
\newblock A class of semi-{L}agrangian approximations for fluids.
\newblock {\em \textit{J. Atmos. Sci.}}, 49(22):2082--2096, 1992.

\bibitem{SK06}
M.~Spiegelman and R.~F. Katz.
\newblock A semi-{L}agrangian {C}rank--{N}icolason algorithm for the numerical
  solution of advection-diffusion problems.
\newblock {\em Geochem. Geophys. Geosyst.}, 7:1--12, 2006.

\bibitem{Staniforth1991}
A.~Staniforth and J.~C\^ot\'e.
\newblock Semi-{L}agrangian integration schemes for atmospheric models--{A}
  review.
\newblock {\em Month. Weather Rev.}, 119(9):2206 -- 2223, 1991.

\bibitem{takahashi_1999}
M.~Takahashi.
\newblock {\em Thermodynamics of One-Dimensional Solvable Models}.
\newblock Cambridge University Press, Cambridge, 1999.

\bibitem{Tok11}
M.~Tokman.
\newblock A new class of exponential propagation iterative methods of
  {R}unge--{K}utta type ({EPIRK}).
\newblock {\em J. Comput. Phys.}, 230:8762--8778, 2011.

\bibitem{https://doi.org/10.48550/arxiv.2208.06614}
R.~S. Watson, S.~A. Simmons, and K.~V. Kheruntsyan.
\newblock Benchmarks of {G}eneralized {H}ydrodynamics for 1d {B}ose gases.
\newblock {\em arXiv:2208.06614v1}, 2022.

\bibitem{XIU2001658}
D.~Xiu and G.~E. Karniadakis.
\newblock A semi-{L}agrangian high-order method for {N}avier--{S}tokes
  equations.
\newblock {\em J. Comput. Phys.}, 172(2):658--684, 2001.

\bibitem{PhysRevLett.19.1312}
C.~N. Yang.
\newblock Some exact results for the many-body problem in one dimension with
  repulsive delta-function interaction.
\newblock {\em Phys. Rev. Lett.}, 19:1312--1315, Dec 1967.

\bibitem{YY1}
C.~N. Yang and C.~P. Yang.
\newblock Thermodynamics of a one-dimensional system of bosons with repulsive
  delta-function interaction.
\newblock {\em J. Math. Phys.}, 10(7):1115--1122, 1969.

\end{thebibliography}
\bibliographystyle{abbrv}


\end{document}